\documentclass[aps,prx,amsmath,notitlepage,twocolumn
]{revtex4-2}

\usepackage{bm} 
\usepackage{graphicx} 
\usepackage[version=4]{mhchem} 
\usepackage{xcolor} 
\usepackage[colorlinks=true,citecolor=blue,linkcolor=blue,urlcolor=blue]{hyperref}%
\usepackage{amsfonts}

\usepackage{float}
\begin{document}


\author{Efe Ilker}
\email{efe.ilker@inserm.fr}
\affiliation{Aix-Marseille Université, INSERM, DyNaMo, Turing Centre for Living Systems, Marseille 13009, France}
\affiliation{Max Planck Institute for the Physics of Complex Systems, 01187 Dresden, Germany}

\author{Kathrin S. Laxhuber}
\affiliation{Max Planck Institute for the Physics of Complex Systems, 01187 Dresden, Germany}
\affiliation{Max Planck School Matter to Life}

\author{Jean-Fran\c{c}ois Joanny}
\email{jean-francois.joanny@college-de-france.fr}
\affiliation{Collège de France, Université Paris Sciences et Lettres,
Matière molle et biophysique, Paris 75231, France}
\affiliation{Institut Curie, Université Paris Sciences et Lettres, Physique des Cellules et Cancer, Paris Cedex 05 74248, France}

\author{Frank Jülicher}
\email{julicher@pks.mpg.de}
\affiliation{Max Planck Institute for the Physics of Complex Systems, 01187 Dresden , Germany}
\affiliation{Center for Systems Biology, 01307 Dresden , Germany}
\affiliation{Cluster of Excellence Physics of Life, Technische Universität Dresden, 01397 Dresden, Germany}

\title{Hydrodynamic theory of chemically active emulsions}

\begin{abstract} 
We present a systematic theory of 
chemically active emulsions in the hydrodynamic limit by constructing a thermodynamically consistent framework in which the equilibrium is broken by chemostating of fuel molecules. For ternary solutions with active chemical reactions, we obtain an effective dynamics of the conserved field dynamics at long length and time scales. The effective dynamics takes into account the broken time reversal symmetry
that manifests itself by the emergence of
gradient terms akin to those of Active Model B+, which is a generic theory of active phase separation. In addition to the active coefficients modifying the interfacial energy coefficient, the theory contains higher order terms in the gradient expansion that are necessary to correctly describe the dynamics of chemically active emulsions, extending thus Active Model B+. We study numerically a Flory-Huggins model with active chemical reactions. Our theory predicts the formation of microphases when the effective interfacial energy coefficient becomes negative. Moreover, including noise, we show the existence of bubbly phase separation. We also identify a new type of phase behavior, a dynamic active filament phase. Finally, we discuss the steady state entropy production rate in the system resulting from the active chemical reactions. We observe that the total entropy production rate increases with the driving chemical potential and exhibits a 
kink-like singularity at the transition to the dynamic active filament phase. Our work shows that the generic behaviors of active phase separation can emerge in chemically active emulsions.
\end{abstract}

\maketitle


\section{Introduction}
Equilibrium emulsions are a metastable state of the phase separation between two liquids (e.g. oil and water) with droplets enriched in 
one component embedded in a continuous phase rich in the other component. The finite droplet size can be maintained if the coarsening 
to a macroscopic phase is extremely slow or if the coalescence between droplets or Ostwald ripening via diffusion are arrested by the addition of a third 
compound, such as surfactants or small colloidal particles as in pickering 
emulsions. Passive emulsions are in general polydisperse in size, even though very subtle procedures have been devised to create monodisperse emulsions \cite{bibette}.

Emulsions can become chemically active if there is a chemical reaction between components that is maintained out of equilibrium by consuming a fuel that
is supplied by reservoirs
\cite{weber_physics_2019}. 
In a biological system such a fuel could be Adenosinetriphosphate (ATP). The hydrolysis of ATP to 
Adenosinediphosphate (ADP) and inorganic phosphate P$_i$ drives many active processes in cells  \cite{julicher1997modeling}. 
The interior of the cell  
hosts numerous chemical processes that are maintained out of equilibrium. Furthermore,
biochemistry is organized in space. For example, biological condensates form that can behave as segregated droplets \cite{banani2017biomolecular,shin2017liquid}.  
The cell interior can therefore be considered as an example of active emulsions \cite{zwicker2025physics}. Chemically active emulsions 
have a  broader range of phase behaviors than passive emulsions, notably spontaneous droplet division \cite{zwicker_growth_2016}, as well as suppression and reversal of the ripening process \cite{zwicker2015suppression,glotzer1995reaction,carati1997chemical,bauermann_theory_2025}.  
Inverse ripening creates an intrinsic characteristic size and can make the structure  more robust to external 
perturbations, which could be relevant to cellular processes. 

Chemically active emulsions are an example of active matter. 
In active matter, various active phase transitions can occur \cite{marchetti2013hydrodynamics}; the two most common examples are flocking \cite{vicsek1995novel,toner1998flocks} 
and motility-induced phase separation \cite{cates2015motility,fily2012athermal}. Their study is  an active field of research and is often based on dynamic field 
theories.  A generic model of active phase separation is the Active Model B+ (AMB+)  introduced in \cite{tjhung2018cluster, nardini2017entropy}. It 
is an active version of the classical model B for the dynamics of phase transitions with a conserved scalar 
order parameter \cite{hohenberg1977rev}. The important physical quantity is the flux of the order parameter, which does not vanish if 
the order parameter is not homogeneous. In the active field theory the flux is expanded in powers of the gradient of the order 
parameter and other spatial derivatives, keeping all terms allowed by symmetry including terms that cannot be obtained 
from a free energy and therefore do not exist in the passive model B \cite{cates2025active}. The AMB+ displays a rich phase 
behavior, including reversed ripening, microphases, foam phases, and bubbly phases \cite{tjhung2018cluster,fausti2021capillary}. The bubbly phase separation, for example, does 
not have any passive equivalent, with a turnover of bubbles inside a macrophase. Some of the phenomenology described by AMB+ has also been observed in theoretical microscopic models. For example, bubble formation within a macrophase appears in mixtures of motile active particles \cite{patch2018curvature, stenhammar2014phase, redner2013reentrant, caporusso2020motility}, while microphases arise in non-motile particles with switching interactions \cite{alston2022intermittent}.

Revealing the connections between the phase behaviors of chemically active emulsions and active field theories remains a key question. In chemical systems, molecular species can be interconverted through chemical reactions. In these reactions, the 
concentrations of substrates and products are not conserved individually, but certain linear combinations of molecular species — such 
as total enzyme or total substrate and product in catalytic reactions — remain conserved due to underlying stoichiometric 
relationships \cite{rao2016nonequilibrium,bauermann2022energy}. These combinations can be described by conserved fields  obeying the continuity equation. Reactions do not 
contribute as source terms in these dynamics, but they contribute to the spatial fluxes describing the time evolution of the conserved fields. Conserved fields generally evolve slowly at long wavelengths, enabling a hydrodynamic description at long time scales by eliminating fast variables \cite{chaikin1995principles}. Thus, the refined problem is to determine how the hydrodynamic description of conserved fields in chemically active emulsions compare with, and differ from, field theories of active phase separation.

In this work, we develop a systematic theory of chemically active emulsions in the hydrodynamic limit.  We study a system with three components, a solvent and two reactants $\rm A$ and $\rm B$ that can be 
converted into one another. The chemical reaction has two parallel pathways, the simple first order reaction from $\rm A$ to 
$\rm B$ and a reaction that consumes an external fuel such as ATP. In a biochemical language, we study an ATPase with two 
isoforms $\rm A$ and $\rm B$ that have different physico-chemical properties and can phase separate between themselves 
or with the solvent.  Starting from a dynamic model of ternary 
solutions with active chemical reactions and Cahn-Hilliard dynamics (Section \ref{sec:ternary}), the hydrodynamic theory is built by expanding the 
dynamic equations at small wave vectors $q$ for $q\ell_d \ll 1$ where the reaction-diffusion length $\ell_d$ is obtained by 
comparing the reaction rate to diffusion (Section \ref{sec:hydro}). We obtain the key results of this work in Section \ref{sec:activeB} for the effective dynamics of a single conserved 
field at long length and time scales. This allows us to determine the parameter region where microphases appear. In particular, leveraging the mapping onto the AMB+ theory facilitates our search on finding exotic phases in chemically active emulsions. We compare the results of the hydrodynamic theory both to AMB+ and to numerical simulations of the full Cahn-Hilliard theory (Section \ref{sec:numerics}). Finally, we discuss dissipation and calculate the entropy production rate for varying chemical activity (Section \ref{sec:epra}).

\section{Ternary solutions with active chemical reactions}\label{sec:ternary}
We consider an incompressible mixture of two solutes $\ce{A}$, $\ce{B}$, and a solvent $\ce{S}$ with volume fractions $\phi_{\ce{A}}$, $\phi_{\ce{B}}$, and $\phi_{\ce{S}}=1-\phi_{\ce{A}}-\phi_{\ce{B}}$. The solutes undergo a chemical reaction $\ce{A}\rightleftharpoons
\ce{B}$. This reaction conserves the density $\psi=\phi_{\ce{A}}+\phi_{\ce{B}}$, whereas the reaction extent $\xi=\phi_{\ce{B}}-\phi_{\ce{A}}$ measures its turnover.  We describe the system state by the variables $\psi$ and $\xi$, and coarse-grain to obtain an effective theory for the slow conserved density $\psi$ by eliminating adiabatically the fast variable $\xi$. The system is governed by the free energy functional
\begin{equation}
\mathcal{F}=\frac{1}{v}\int d{\bf r} \left(f(\psi,\xi)+\frac{\kappa}{2}(\nabla\psi)^2
\right)  
\label{eq:freefunctional}
\end{equation}
which sets the chemical potentials through the functional derivatives $\mu_\psi(\psi,\xi)=\delta 
\mathcal{F}/\delta \psi$ and $\mu_\xi(\psi,\xi) = \delta \mathcal{F}/\delta \xi$ where $\kappa$ is the interfacial energy coefficient and $v$ is the volume of molecules A and B which we consider to be equal sized. Throughout this paper, we use a convention that $\mathcal{F}$ and chemical potentials are given in units of $k_B T$ where $k_B$ is Boltzmann constant and $T$ is the local temperature.

The spatiotemporal dynamics of $\psi$ follows a conservation law and evolves through transport 
fluxes. The reaction extent $\xi$ evolves through transport fluxes and reaction fluxes. 
Accordingly, we write 
\begin{eqnarray}
\frac{\partial \psi}{\partial t} &=& -\nabla \cdot \bm{j}_\psi\nonumber \\
\frac{\partial \xi}{\partial t} &=& -\nabla \cdot \bm{j}_\xi  + 2r \quad , \label{eq:dynamics}
\end{eqnarray}
where $r$ denotes the reaction flux and the transport fluxes $\bm{j}_{\psi}, \bm{j}_{\xi}$ 
are given in Cahn-Hilliard form by
\begin{eqnarray}
\bm{j}_\psi= -\Lambda_{\psi\psi}\nabla \mu_\psi -
\Lambda_{\psi\xi}\nabla \mu_\xi \quad \nonumber \\
\bm{j}_\xi = -\Lambda_{\xi\psi}\nabla \mu_\psi -
\Lambda_{\xi\xi}\nabla \mu_\xi \quad,  \label{eq:js}
\end{eqnarray}
where $\Lambda_{ij}$ denote the mobility matrix, $i,j\in\{\psi, \xi \}$, which in general depends on composition $(\psi,\xi)$. $\Lambda_{ij}$ is a symmetric positive definite matrix to ensure the second law of thermodynamics and the Onsager reciprocity relations. The transformations of mobilities and chemical potentials from variables $\phi_A$ and $\phi_B$ to variables $\psi$ and $\xi$ are given in Appendix \ref{app:lineartransform}.

The rate $r$ of the chemical reaction 
$\ce{A}\rightleftharpoons \ce{B}$ is the sum of two  parallel processes: a 
passive reaction with rates $k_1^\pm$ and an active process with rates $k_2^\pm$, where $(+)$ denotes the
forward reaction $\ce{A} \rightarrow \ce{B}$ and $(-)$ the backwardreaction $\ce{A} \leftarrow \ce{B}$. The passive 
process satisfies a detailed balance relation $k_1^+/k_1^-=e^{-2\mu_\xi}$. The active process is driven by the chemical free energy $\Delta\mu$ of a fuel, which obeys the relation $k_2^+/k_2^-=e^{-2{\mu_\xi}+\Delta\mu}$.
The Gibbs free energies that drive the chemical reactions are respectively $2 \mu_{\xi}$ and $2 \mu_{\xi}-\Delta \mu$ for passive and active reactions and their signs determine the net directions of reaction fluxes. We hence write $r=k_1^+(1-e^{2\mu_\xi})+k_2^+(1-e^{2\mu_\xi-\Delta\mu})$.
We can write the reaction flux $r$ as
\begin{equation}
 r=k(1-(1-\alpha)e^{2\mu_\xi})\label{eq:r} \quad ,
\end{equation}
where $k=k_1^++k_2^+$ and
\begin{equation}
\alpha=\frac{k_2^+}{k_1^++k_2^+}(1-e^{-\Delta\mu}) \quad . \label{eq:alpha}
\end{equation}
The activity coefficient $\alpha(\psi, \xi)$ depends on the composition and measures the non-equilibrium character of the reactions. It has values $0\leq\alpha\leq1$ for $\Delta \mu \geq 0$. For $\Delta\mu=0$ we have $\alpha=0$, and the system relaxes to an equilibrium steady state, which minimizes the free energy $\mathcal{F}$. For a general $\alpha(\psi,\xi)$, we show in the following how the conserved density $\psi$ is driven out of equilibrium.

\section{Hydrodynamic expansion at long length and time scales} \label{sec:hydro}
At large length scales $L$, the conserved density $\psi$ relaxes slowly over a time scale $L^2/\bar{D}$, 
where $\bar{D}$ is a characteristic diffusion coefficient. If this time is long compared to the 
characteristic time scale $\bar{k}^{-1}$ of the chemical reactions, i.e., $L^2/\bar{D} \gg 1/\bar{k}$, 
the non-conserved reaction extent $\xi$ relaxes instantaneously to a value determined by the slow field 
$\psi$. In order to determine the resulting effective dynamics of $\psi$, we consider a hydrodynamic limit.
We rescale time as $t=\tau L^2/\bar{D}$ and lengths as $x=\rho L$ and define the small parameter $\epsilon=\bar{D}/(L^2 
\bar{k})$. In the new variables $(\rho,\tau)$, the equations of motion are dimensionless and read
\begin{eqnarray}
\epsilon\frac{\partial\xi}{\partial \tau}&=&-\frac{\epsilon}{\bar{D}}\nabla_{\rho} \cdot \bm{j}_\xi 
+\frac{2r}{\bar k} \label{eq:xieps} \\
\frac{\partial\psi}{\partial \tau}&=&-\frac{1}{\bar{D}}\nabla_{\rho}\cdot \bm{j}_\psi \quad ,\label{eq:psips}
\end{eqnarray}
where derivatives $\nabla_{\rho}$ and fluxes $\bm{j}$ are defined with respect to $\rho$.  The hydrodynamic limit is obtained by expanding Eq.~\eqref{eq:xieps} to the first order in $\epsilon$ and by eliminating the reaction extent $\xi$, which is a fast variable, to obtain an effective equation for the slow variable 
$\psi$. We therefore look for expansions of the reaction extent and the associated chemical potential of the form  $\xi=\xi_0+\epsilon \xi_1+\dots$ and $\mu_\xi=\mu_\xi^{(0)}+\epsilon \mu_\xi^{(1)}+\dots$.  To order $\epsilon =0$, at large scales, chemical reactions are stationary for a given density $\psi$ and Eq.~\eqref{eq:xieps} leads to a balanced reaction flux $r=0$. Therefore using \eqref{eq:r} we have
\begin{equation}
\mu_\xi^{(0)}(\psi,\xi_0)=-\frac{1}{2}\ln(1-\alpha) \quad . \label{eq:muxi0}
\end{equation}
For given $\alpha$, this implies a dependence $\xi_0=g(\psi)$ on $\psi$. The function $g(\psi)$ can be calculated for a 
specific system given its free energy from Eq.~\eqref{eq:freefunctional} and its chemical driving $\alpha(\psi,\xi)$. Its derivative is given by $g'= d g/d\psi$ with 
\begin{equation}
    g'=\frac{\frac{1}{2}\frac{\partial \alpha}{\partial \psi}/(1-\alpha) -  \chi^{-1}_{\xi \psi}}{\chi^{-1}_{\xi \xi}-\frac{1}{2}\frac{\partial \alpha}{\partial \xi} /(1-\alpha)}\label{eq:gprime}\quad ,
\end{equation} where the osmotic compressibility matrix is defined by $\chi^{-1}_{ij}=
\partial \mu_i/\partial\phi_j$ and $i =(\psi, \xi)$.

To first order in $\epsilon$, Eq.~\eqref{eq:xieps} leads to 
\begin{equation}
g'\frac{\partial \psi}{\partial \tau}=-\frac{1}{\bar{D}}\nabla_{\rho}\cdot \bm{j}_\xi^{(0)}-\frac{4k}{\bar k}\mu_\xi^{(1)} \label{eq:Oepsilon}
\end{equation}
with $\bm{j}_\xi^{(0)} = -\Lambda_{\xi\psi}\nabla_{\rho} \mu_\psi^{(0)}-\Lambda_{\xi\xi}\nabla_{\rho} \mu_{\xi}^{(0)}$. 
The superscript $0$ denotes an evaluation at order $\epsilon=0$ with $(\psi,\xi)= (\psi,g(\psi))$. 

This approach allows us to systematically eliminate the reaction extent $\xi$ and to obtain an effective equation for the conserved density in the hydrodynamic limit.  We refer to appendix~\ref{app:expansion} for 
details. Briefly, we use Eq. (\ref{eq:psips}) to replace
$\partial \psi/\partial \tau=-\bar{D}^{-1}\nabla_{\rho}\cdot{\bm j}_\psi^{(0)}$ in Eq.~\eqref{eq:Oepsilon}, where ${\bm j}_\psi^{(0)}=-
\Lambda_{\psi\psi}\nabla_{\rho}\mu_\psi^{(0)}-\Lambda_{\psi\xi}\nabla_{\rho}\mu_\xi^{(0)}$  and determine $\mu_{\xi}^{(1)}$. Given $\mu_{\xi}^{(1)}$, we 
then determine the correction $\epsilon\xi_1$ to $\xi(\psi)$, and finally take these corrections into account in all $\xi$-dependent 
quantities up to a given order, while dropping higher order terms. 

Returning to the original time and space variables, the effective dynamics of the conserved density $\psi$ up to first order in $\epsilon$ are then given by:
\begin{eqnarray}
\frac{\partial \psi}{\partial t} &=& -\nabla \cdot \bm{j}  \quad , \label{eq:dpsieff}\\ 
\bm{j} &=& -\Lambda_{\psi\psi}\left(\nabla(\mu_{\psi}^{(0)}+\epsilon\mu_{\psi}^{(1)})+s\nabla
(\mu_{\xi}^{(0)}+\epsilon\mu_{\xi}^{(1)}) \right)\label{eq:jpsi}  \ ,
\end{eqnarray}
where $s=\Lambda_{\psi\xi}/\Lambda_{\psi\psi}$, which we also expand to first order.
We discuss the form of the current $\bm{j}$ in the next section, mapping the coefficients of the two component model with chemical reactions to parameters characterizing the AMB+.

\section{Effective coarse-grained theory
} \label{sec:activeB}
In the hydrodynamic limit, we obtain the  dynamic equation (\ref{eq:jpsi}) for the conserved slow variable $\psi$ of the chemically active ternary solution: 
\begin{eqnarray}
\hat M^{-1}\cdot\bm{j}= &-& \nabla (\mu_{\rm eff} +\lambda_{\rm eff} (\nabla\psi)^2+\nu_{\rm eff} \nabla^4\psi )\nonumber \\
&+&\zeta_{\rm eff}(\nabla^2\psi)\nabla\psi + \eta_{\rm eff} (\nabla \psi)^3\quad  \label{eq:jresult}
\end{eqnarray}
Here, the operator $\hat M$ is a mobility and the effective chemical potential $\mu_{\rm eff}$ can be obtained from an effective free energy $\mu_{\rm eff}=\delta \mathcal{F}_{\rm eff}/\delta \psi$. It reads
\begin{equation}
\mu_{\rm eff}=\mu_{\psi}(\psi,g(\psi))+\int_0^\psi \mathrm{d}\psi'
s\beta - \kappa_{\rm eff}\nabla^2 \psi-\frac{K'}{2}(\nabla\psi)^2 \quad ,
\label{eq:mueff}
\end{equation}
where  
\begin{eqnarray}
\kappa_{\rm eff}&=&\kappa+K\label{eq:kappaeff}\\
K&=&\frac{\Gamma\beta}{4k}
\left (g'-\gamma h-s\right )\quad,\label{eq:l}
\end{eqnarray}
renormalizing the coefficient $\kappa$ of the gradient term in the free energy. Here and below, primes denote total derivatives with respect to $\psi$,
i.e., $F'\equiv\partial F/\partial \psi + g'\partial F/\partial\xi$ for a function $F(\psi,\xi)$. Moreover,
we have defined  
\begin{eqnarray}
\gamma &\equiv&\frac{\partial \mu_\xi^{(0)}}{\partial \psi} 
= \frac{1}{2}\frac{\partial\alpha/\partial\psi}{1-\alpha} \quad ,\\
\beta &\equiv& \frac{ d \mu_\xi^{(0)}}{d \psi } = \frac{1}{2}\frac{\alpha'}{1-\alpha} \quad ,
\end{eqnarray}
and 
\begin{eqnarray}
h&=&\frac{1}{\chi^{-1}_{\xi \xi}-\frac{1}{2}\frac{\partial \alpha}{\partial \xi} /(1-\alpha)}
\quad ,\\
\Gamma&=&\Lambda_{\xi\xi}-\Lambda_{\psi\xi}^2/\Lambda_{\psi\psi} \quad .
\end{eqnarray}
$\Lambda_{\psi\psi}\Gamma$ is the determinant of the mobility matrix, and thus $\Gamma>0$.

On the one hand, a system for which  $\mu_{\rm eff}$ alone does not vanish can be described by an effective equilibrium theory but with renormalized  coefficients due to the non-equilibrium chemical reaction. 
On the other hand, the active nature of the system is captured by the other contributions to the current characterized by the following coefficients:
\begin{eqnarray}
\zeta_{\rm eff}&=& \frac{\Gamma\beta}{4k}\left(s'-s_{\xi}\beta h\right)\quad , \label{eq:zeta} \\  
\lambda_{\rm eff}&=& \frac{K'}{2}-\frac{(\Gamma\beta)'}{4k} (g'-\gamma h-s) \quad , \label{eq:lambda} \\
\eta_{\rm eff} &=& \frac{(\Gamma\beta)'}{4k} \left(s'-s_{\xi}\beta h \right) \label{eq:eta} \quad ,\\
\nonumber\
\end{eqnarray}
where we have defined $s_{\xi}=\partial s/\partial \xi$. The  quantities characterizing the mobility matrix $s$, 
$\Gamma$, and  $s_{\xi}$
are all evaluated at $\xi=g(\psi)$. The mobility operator in Eq.~\eqref{eq:jresult} is given by
\begin{eqnarray}
\hat M = &&\Lambda_{\psi \psi} \Bigl( 
1 - \nabla (g'-\gamma h)\hat A 
+ s \nabla \hat A \nonumber \\
&& +  \nabla \mu_{\xi}^{(0)} s_{\xi} h \hat A
- \mu_{\xi}^{(0)} \nabla s_{\xi} h \hat A \Bigr)\label{eq:M}
\end{eqnarray}
where we have defined the operator
$\hat A=(\nabla \Lambda_{\psi\xi}-
g'\nabla \Lambda_{\psi\psi})/4k$.
Note that in the hydrodynamic limit, $\hat M$ remains always positive definite because all gradient terms are small compared to $1$. For the detailed derivations of Eqs. \eqref{eq:mueff}-\eqref{eq:M}, see Appendix \ref{app:expansion}.

Finally, in the case where the effective interfacial coefficient becomes negative due to active chemical reactions, i.e., when $\kappa_{\rm eff}<0$, the term  $\nu_{\rm eff}\nabla^4\psi$ is required for linear stability whose coefficient is given by:
\begin{equation}
\nu_{\rm eff}=-\frac{\Gamma}{4k}K \quad .   \label{eq:nueff} 
\end{equation} 
The details of its derivation are given in Appendix \ref{app:q6}. Since $\Gamma, k>0$, $\nu_{\rm eff}$ has the opposite sign with $K$ which grants a stabilizing contribution. For simplicity, we neglected the other terms at the same gradient order in writing Eq.~\eqref{eq:jresult} that do not contribute to linear stability. These terms can be calculated by extending our systematic hydrodynamic expansion (Section \ref{sec:hydro} and Appendix \ref{app:expansion}). As we discuss in the next section, we observe that \eqref{eq:jresult} is sufficient to capture qualitative features in chemically active emulsions in the hydrodynamic limit.

The coefficients of the gradient terms of  Eq.~\eqref{eq:jresult}, given by \eqref{eq:eta} mirror those defined in the AMB+ \cite{tjhung2018cluster}. However, the three coefficients are not constant and depend on the total concentration $\psi$ of reactants. We also obtain additional terms $\eta_{\rm eff} (\nabla\psi)^3$ and $-\nabla \left(\nu_{\rm eff} \nabla^4\psi\right)$. The first arises at the same gradient order which is also recently discovered in coarse-graining of thermal quorum-sensing active particles \cite{burekovic2026active}. The latter is particularly an important correction for chemically active emulsions when $\kappa_{\rm eff}$ is small or negative, thus extends the scope of AMB+. This term can be viewed as an integrable term for an effective free energy description, however, it would lead to compensating contributions to active terms at the same gradient order which we did not explicitly calculated. We discuss this further in the model studied in the next section. Overall, we have provided  explicit expressions for the coefficients $\lambda_{\rm eff}$,  $\zeta_{\rm eff}$ and $\eta_{\rm eff}$ of the active terms and relate them as well as  $\mu_{\rm eff}$ and $\nu_{\rm eff}$ to the chemical properties and driving of the underlying ternary solution. 

We make the following remarks: i) The magnitudes of all the gradient coefficients depend on  the ratio $\Gamma/k$ that is related to the typical reaction-diffusion length in the system.  ii) For a non-driven passive system with $\alpha = 0$, the coefficients become  $K=0$ and $\zeta_{\rm eff}=\lambda_{\rm eff}=\eta_{\rm eff}=0$.  We then recover equilibrium dynamics. iii) An effective equilibrium is also recovered when $\alpha\neq 0$ but is constant and independent of $\psi$ and $\xi$ with $K=\zeta_{\rm eff}=\lambda_{\rm eff}=\eta_{\rm eff}=0$. This limit amounts to  shifting the equilibrium Gibbs free energy of chemical reactions by a constant value everywhere. In this case, the system would still break  real thermodynamic equilibrium with $\alpha\neq 0$ and, while $\alpha$ appears as an extra parameter controlling the phase separation without changing the phenomenology. A similar case was shown in other non-equilibrium systems such as in the dilute limit of mixtures of particles having two different temperatures \cite{ilker2020phase}. iv) For constant $s=\Lambda_{\psi\xi}/\Lambda_{\psi\psi}$ the effective dynamics becomes that of Active Model B with $\zeta_{\rm eff}=\eta_{\rm eff}=0$ while $\lambda_{\rm eff}$ can be non-zero.

Thus, a requirement for the new types of phases predicted by AMB+ is that the active coefficient $\alpha$ is not constant, allowing the broken detailed balance of reactions to be transmitted to the diffusive flux of $\psi$. An additional requirement for the ternary solutions described in Section \ref{sec:ternary} is that $s$ should not be constant. Note that the latter is a less strict condition for a system governed by a  energy functional that includes additional interfacial terms, e.g., $\nabla \psi\nabla\xi$, and $(\nabla \xi)^2$. For brevity, we did not include these terms in Eq.~\eqref{eq:freefunctional}, as they do not affect the phenomenology and only rescale the effective coefficients. Similarly, the free energy functional can contain a passive contribution proportional to $(\nabla^2 \psi)^2$ that would be added to $\nu_{\rm eff}$. 

Given a specific system with certain free energy, mobilities, chemical rates, and driving $\Delta\mu$, one can calculate first $\xi_0=g(\psi)$ from Eq.~\eqref{eq:muxi0} and then obtain the coefficients from Eqs.~\eqref{eq:eta} as well as the effective chemical potential and $\nu_{\rm eff}$ from Eqs.~\eqref{eq:mueff},~\eqref{eq:l}, and \eqref{eq:nueff}. In the following section,  we numerically study a specific example of ternary system by choosing a free energy density $f(\psi, \xi)$ and an activity parameter $\alpha(\psi, \xi)$.

\section{Active phases} \label{sec:numerics}
In this section, we study an example of ternary solution following the dynamics described in Section \ref{sec:ternary} by introducing a model for the free energy density $f$ and the activity parameter $\alpha$. 
\subsection{Flory-Huggins model with active chemical reactions} \label{sec:model}

We consider that the thermodynamics of the system is governed by a Flory-Huggins free energy in the presence of chemical reactions and with a single a repulsive interaction between solute B and  solvent. We write the bulk part of the free energy as:
\begin{eqnarray}
    f(\phi_{\rm A},\phi_{\rm B})&=&\phi_{\rm A} \ln \phi_{\rm A}+\phi_{\rm B} \ln \phi_{\rm B}+\phi_{\rm S} \ln \phi_{\rm S}\nonumber \\&+&\chi \phi_{\rm B}
   \phi_{\rm S}+\Delta(\phi_{\rm B}-\phi_{\rm A}) \quad. \label{eq:ffh}
\end{eqnarray}
Here the first three terms are the entropic mixing contributions of the two solutes and the solvent with densities $\phi_{\rm A}$, $\phi_{\rm B}$, and $\phi_{\rm S}=1-\phi_{\rm A}-\phi_{\rm B}$ respectively. The fourth term is the interaction between solute B and the solvent, and the last term  term controls the equilibrium value of the reaction extent: $\Delta<0$ favors the forward reaction $\ce{A <=> B}$ and vice versa. The chemical potentials are obtained by expressing the free energy functional \eqref{eq:freefunctional} as a function of the variables  $\psi,\xi$ by inserting \eqref{eq:ffh} with substitutions $\phi_{\ce{A}}=(\psi-\xi)/2$, $\phi_{\ce{B}}=(\psi+\xi)/2$  and taking the functional derivatives: $\mu_\psi(\psi,\xi)=\delta \mathcal{F}/\delta \psi$ and $\mu_\xi(\psi,\xi) = \delta \mathcal{F}/\delta \xi$:
\begin{eqnarray}
\mu_{\psi}&=&\frac{1}{2}\ln\left(\frac{\psi^2-\xi^2}{4(1-
\psi)^2}\right)+\frac{\chi}{2}\left(1-2\psi-\xi \right) -\kappa\nabla^2\psi\quad ,\nonumber\\
\mu_{\xi}&=&\frac{1}{2}\ln \left(\frac{\psi+\xi}{\psi-\xi
}\right) +\frac{\chi}{2}\left(1-\psi \right) +\Delta \quad .\label{eq:mu_fh}
\end{eqnarray}

We  model the kinetics of the non-equilibrium chemical reactions by an active coefficient $\alpha$ and rate $k$:
\begin{eqnarray}
    \alpha&=&(1-e^{
\Theta(\psi
)})(1-e^{-\Delta \mu}), 
\nonumber\\ \Theta(\psi)&=&-b(1-\psi)^2 + a(1-\psi)^3\ ,\label{eq:modelalpha}\\
k&=&k_m \phi_{\rm A}=k_m(\psi-\xi)/2 \label{eq:kfunc}
\end{eqnarray}
where $k_m$ has a constant value. Our choice of the activity coefficient is motivated by the fact that it allows for an analytical solution of \eqref{eq:muxi0} for $g(\psi)$ when $\Delta \mu\gg1 $ and in the remainder we first consider this limit. However, we consider a wider range of  $\Delta \mu$ in  numerical solutions. Considering  $\Delta \mu\geq 0$ imposes $b\geq a$ according to Eq. \eqref{eq:modelalpha}, so that   $0\leq\alpha\leq1$. 

Finally, we set the mobility matrix for the  molecular concentrations $(\phi_{\rm A},\phi_{\rm B})$ as $\tilde{\Lambda}_{mn}=\Lambda_0 \phi_{m}(\delta_{mn}-\phi_{n})$ where $\delta_{mn}$ is the Kronecker delta function and $m,n \in \{A,B \}$ with $\Lambda_0>0$  ensuring positive definiteness. In the coordinate frame of the density fields $\psi, \xi$, we get using \eqref{eq:trmob}
\begin{eqnarray}
    \Lambda_{\psi\psi}&=&{\Lambda_0}(1-\psi)\psi\quad ,\nonumber\\
    \Lambda_{\psi\xi}&=&\Lambda_{\xi\psi}={\Lambda_0}(1-\psi)\xi\quad,\nonumber\\
    \Lambda_{\xi\xi}&=&{\Lambda_0}(\psi-\xi^2) \quad . \label{eq:mobs}
\end{eqnarray}

\begin{figure*}[t]
    \centering
\includegraphics[width=\textwidth]{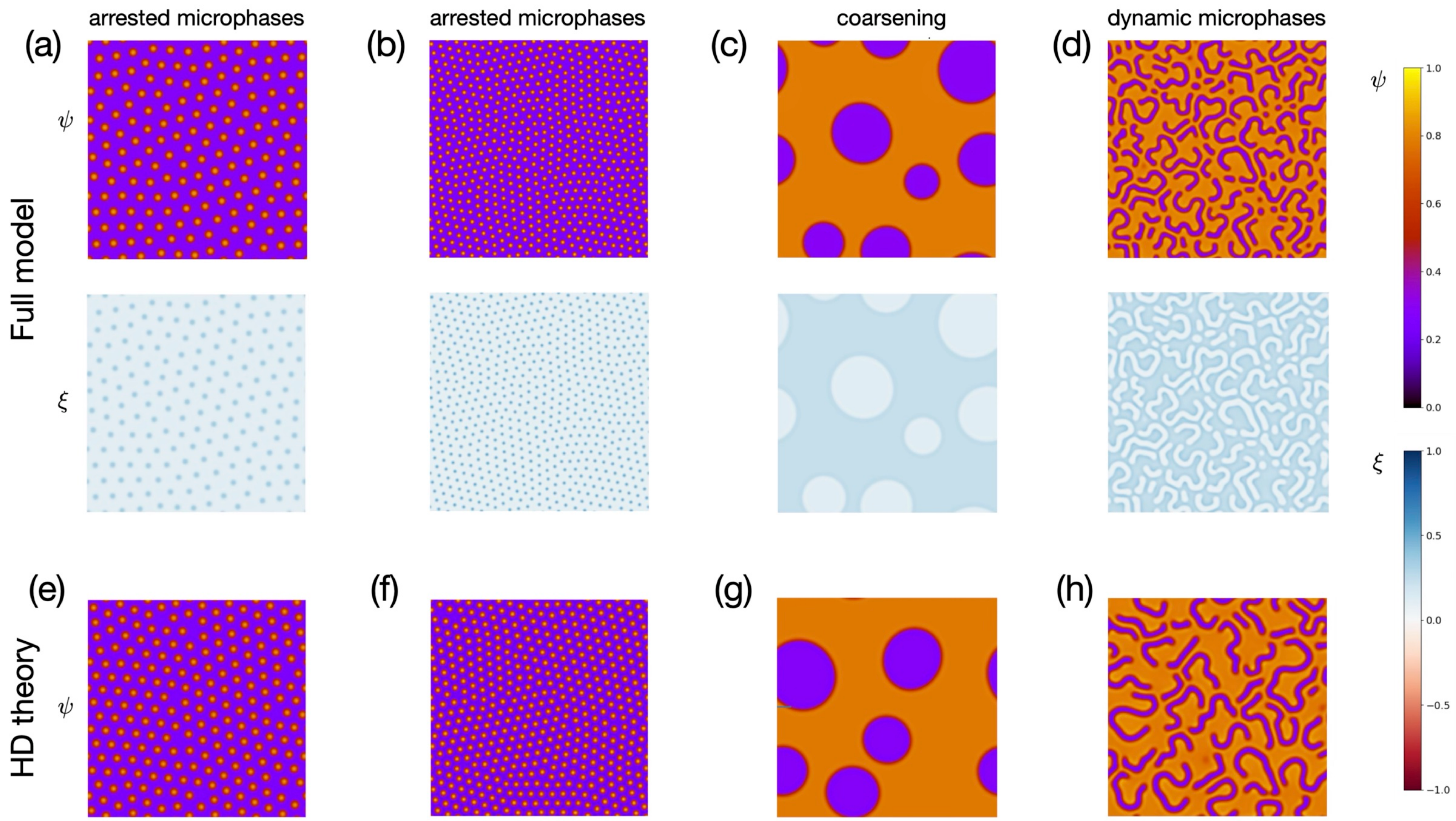}
    \caption{ Snapshots at long times ($k_mt>5\times 10^5$) after initiating from a mixed state. For the full model, we take $k_m=0.558$, $\Lambda_0=1.0$ (other parameters in Section \ref{sec:params}) and show the results for conserved field $\psi$ and reaction extent $\xi$ using (a) $\langle \psi\rangle=0.365$, $\kappa=1.8$, (b)  $\langle \psi\rangle=0.365$, $\kappa=1.0$, (c)  $\langle \psi\rangle=0.65$, $\kappa=1.8$, (d)  $\langle \psi\rangle=0.65$, $\kappa=1.0$. We also show the snapshots from the results using our hdyrodynamic theory (HD theory) with calculated effective coefficients $\lambda_{\rm eff}=-1.04$, $\zeta_{\rm eff}=-3.91$, $\eta_{\rm eff}=-4.35$,  and $\nu_{\rm eff}=2.33$. Results are shown for (e) $\langle \psi\rangle=0.365$, $\kappa_{\rm eff}=-0.23$, (b)  $\langle \psi\rangle=0.365$, $\kappa_{\rm eff}=-1.03$, (c)  $\langle \psi\rangle=0.65$, $\kappa_{\rm eff}=-0.23$, (d)  $\langle \psi\rangle=0.65$, $\kappa_{\rm eff}=-1.03$.   }
    \label{fig:phasediagram}
\end{figure*}

\subsection{Effective coefficients of the hydrodynamic theory} \label{sec:effcoeff}
When $\Delta \mu\gg1$, at lowest order, the reaction extent is calculated from Eqns.~\eqref{eq:muxi0}, \eqref{eq:mu_fh} and \eqref{eq:modelalpha} leading to
$g(\psi)= \psi(e^{\theta(\psi)}-1)/(e^{\theta(\psi)}+1)$ where $\theta(\psi)=-2\Delta-\chi(1-\psi)-\Theta(\psi)$.
We also get $\beta (\psi)=-\Theta'(\psi)/2$, $\gamma (\psi)=\beta(\psi)$ and $h(\psi)=(\psi^2-g(\psi)^2)/\psi$.

Using the form of mobility matrix \eqref{eq:mobs}, we obtain 
$s=\xi/{\psi}$, $s_{\xi}=1/\psi$ and $\Gamma={\Lambda_0}(\psi^2-\xi^2)/\psi$. 
With these expressions, we are able to calculate the effective gradient coefficients using Eqs.~\eqref{eq:l},~\eqref{eq:nueff}, and \eqref{eq:eta}:
\begin{eqnarray}
K&=&-\left(\frac{\Lambda_0}{4k}\right)\frac{4e^{2\theta(\psi)}\psi^2\chi\Theta'(\psi)}{(e^{\theta(\psi)}+1)^4} \ , \label{eq:lfh}\\
\zeta_{\rm eff}&=&\frac{K}{\psi} \quad , \label{eq:zetafh} \\  
\lambda_{\rm eff}&=& \frac{K'}{2}-\frac{(\Gamma\beta)'}{\Gamma\beta}K \quad , \label{eq:lambdafh} \\
\eta_{\rm eff} &=& \frac{(\Gamma\beta)'}{\Gamma\beta} \zeta_{\rm eff}\quad ,\label{eq:etafh} \\ 
\nu_{\rm eff} &=& -\frac{\Gamma}{4k}K \quad ,\label{eq:nuefffh} 
\end{eqnarray}
where $\Gamma\beta=-2\Lambda_0e^{\theta(\psi)
}\psi\Theta'(\psi)/(e^{\theta}+1)^2$.
We notice that the sign  of $K$ and $\zeta_{\rm eff}$ is determined by $\Theta'(\psi)=2b(1-\psi)-3a(1-\psi)^2$ when $\chi>0$ since $\Lambda_0, k$ are both positive.

\subsection{Parameter values}\label{sec:params}

Throughout the paper, we fix the following parameters. First, in free energy density given in \eqref{eq:ffh}, we take $\chi=3.5$, $\Delta=-0.55$. With this choice, the phase separation becomes coupled with chemical reactions.  Second, in the activity parameter $\alpha$ given in \eqref{eq:modelalpha}, we take $a=3.0$, $b=6.56$. As a result, $\alpha$ is monotonically decreasing function of $\psi$ and $\Theta'(\psi)<0$, which sets directly the sign of effective coefficients $K<0$ and $\zeta_{\rm eff}<0$ as we discussed in the previous section. With these parameters fixed, the active phase behavior is controlled by the ratio of interface and reaction-diffusion length scales $\sqrt{\kappa/ (\Lambda_0/k_m)}$ and the driving chemical potential $\Delta \mu$. 

\subsection{Arrested and dynamic microphases}\label{sec:5d}

We now study the phase behavior in this model with numerical solutions in two dimensional systems. The numerical methods are described in Appendix \ref{app:numsol}. 

We first consider the emergence of microphase separation when $\kappa_{\rm eff}<0$. A finite wavelength is stabilized by the term  $\nu_{\rm eff}$ given by Eq.~\eqref{eq:nueff} which becomes positive for $K<0$. In order to explore the phenomenology, we may neglect the variation of $\nu_{\rm eff}$ around its value at the interface  $\nu_{\rm eff}(\psi^{\rm int})$ and consider it as a constant. Then for $\zeta_{\rm eff}=\lambda_{\rm eff}=\eta_{\rm eff}=0$ the system dynamics is governed by a chemical potential that is integrable to a free energy of the form $F=\int d{\bf r} \left(f(\psi,\xi)+\frac{\kappa_{\rm eff}}{2}(\nabla\psi)^2+\frac{\nu_{\rm eff}}{2}(\nabla^2\psi)^2\right)$. Such equilibrium-like systems form stationary periodic mesophases for $\kappa_{\rm eff}<0$, $\nu_{\rm eff}>0$ \cite{cross1993pattern,elder2002modeling}. These would yield fixed size dense droplets in a dilute background for low $\langle \psi \rangle$ values and fixed size bubbles (dilute droplets) in a dense background for high $\langle \psi \rangle$ values. However, the AMB+ terms $\lambda_{\rm eff}$ and $\zeta_{\rm eff}$  break the symmetry between dilute and dense droplet formation.

To illustrate an example, we consider a system with $\Delta \mu=20.0$ for which Eqs. \eqref{eq:lfh}-\eqref{eq:nuefffh} provide an excellent approximation. Although all effective coefficients have $\psi$ dependence, for simplicity we restrict our attention to their value at the interface, which should provide a good approximation. We determine the interface value $\psi^{\rm int}=(\psi_{\rm I}+\psi_{\rm II})/2$ with coexisting phase values $\psi_{\rm I}$ and $\psi_{\rm II}$ of the free energy functional $\mathcal{F}_{\rm eff}$ defined with \eqref{eq:mueff} using the model described in section~\ref{sec:model}. This gives $\psi^{\rm int}=0.516$. Moreover, we set $k_m=0.558$ and $\Lambda_0=1.0$. As a result, the interface value of the coefficients are $K=-2.03$, $\lambda_{\rm eff}=-1.04$, $\zeta_{\rm eff}=-3.91$, $\eta_{\rm eff}=-4.35$,  and $\nu_{\rm eff}=2.33$. In Fig. \ref{fig:phasediagram}, we show the phase behavior in the region where $\kappa_{\rm eff} = \kappa + K<0$ for different $\kappa$ and space-averaged density $\langle \psi \rangle$. In the low density region when $\langle \psi \rangle=0.365$, we observe arrested microphases with dense droplets having regular arrangements for $\kappa=1.8$ (Fig. \ref{fig:phasediagram}(a)) and $\kappa=1.0$ (Fig. \ref{fig:phasediagram}(b)), and the typical distance between the dense droplets decreases with decreased $\kappa$. By contrast, for $\kappa=1.8$ and $\langle \psi \rangle=0.65$, we observe coarsening dynamics toward a macro droplet, as shown in Fig. \ref{fig:phasediagram}(c), unlike for $\langle \psi \rangle=0.365$. Interestingly, for $\kappa=1.0$ and $\langle \psi \rangle=0.65$, the system exhibits dynamic (chaotic) filament-like phases as shown in the snapshot in Fig. \ref{fig:phasediagram}(d)). Here, the filaments continuously move and exhibit elongation and fission dynamics. Movies will be available with publication. In the appendix, we also show the snapshots in terms of $\phi_{A}$ and $\phi_B$ where both volume fractions follow the observed patterns (see Fig.~\ref{fig:ab}).

We compare the results of the full model with our hydrodynamic theory by solving \eqref{eq:dpsieff}, \eqref{eq:jresult} using the interface value of the coefficients obtained in the above paragraph. The snapshots of $\psi$ at long times are shown in Fig.  \ref{fig:phasediagram} (e), (f), (g), (h) below the corresponding full model results. These show excellent agreement with the full model in capturing the phase behavior. Moreover, we compare the characteristic length scales particularly for the arrested microphases shown in Fig. \ref{fig:phasediagram} (a), (b) and (e), (f). The typical distance between dense droplets is $\ell=2\pi/q^*$ where $q^*$ is the mean wave number  obtained by using the structure factor (see Appendix \ref{app:structure}). The calculations for the full model result $q^*=0.386$ and $q^*=0.699$ for $\kappa=1.8$ and $\kappa=1.0$ respectively. The values obtained in the hydrodynamic theory snapshots are respectively $q^*=0.403$ and $q^*=0.5764$. Thus, for $\kappa=1.8$ the discrepancy is within $5\%$ while for $\kappa=1.0$ it is within $17\%$. Although small, the increasing discrepancy is likely due to the increase in $q^*$, on which the hydrodynamic expansion is based. Overall, the hydrodynamic theory successfully describes the emerging phases.  

\begin{figure}[h]
    \centering
    \includegraphics[width=0.97\columnwidth]{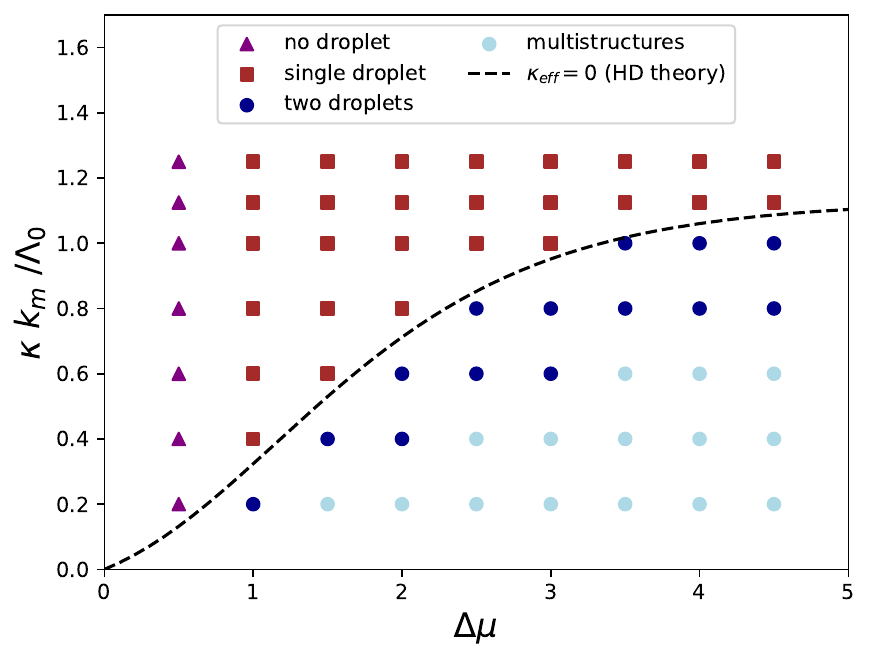}
   \caption{Ripening diagram of the full model with varying $\kappa/ (\Lambda_0/k_m)$ and $\Delta \mu$ for $\langle \psi \rangle=0.365$ (low density regime). Starting from a system of two droplets with unequal sizes, the system reaches one of four states at long times: well-mixed state so no droplet (purple triangels), a single stable droplet via Ostwald ripening (red squares), two droplets via reverse Ostwald ripening (dark blue dots), or multiple microstructures (light blue dots). The dashed line marks the $\kappa_{\rm eff}=0$ in the hydrodynamic theory calculated using Eq.\eqref{eq:kappaeff} with \eqref{eq:lfh} at $\psi^{\rm int}=0.516$ which corresponds to the interface value in the large $\Delta \mu$ limit (see text for details). Thus, the hydrodynamic theory agrees well with having microphases for $\kappa_{\rm eff}<0$.}
    \label{fig:ripening}
\end{figure}

\begin{figure*}[t]
    \centering   \includegraphics[width=\textwidth]{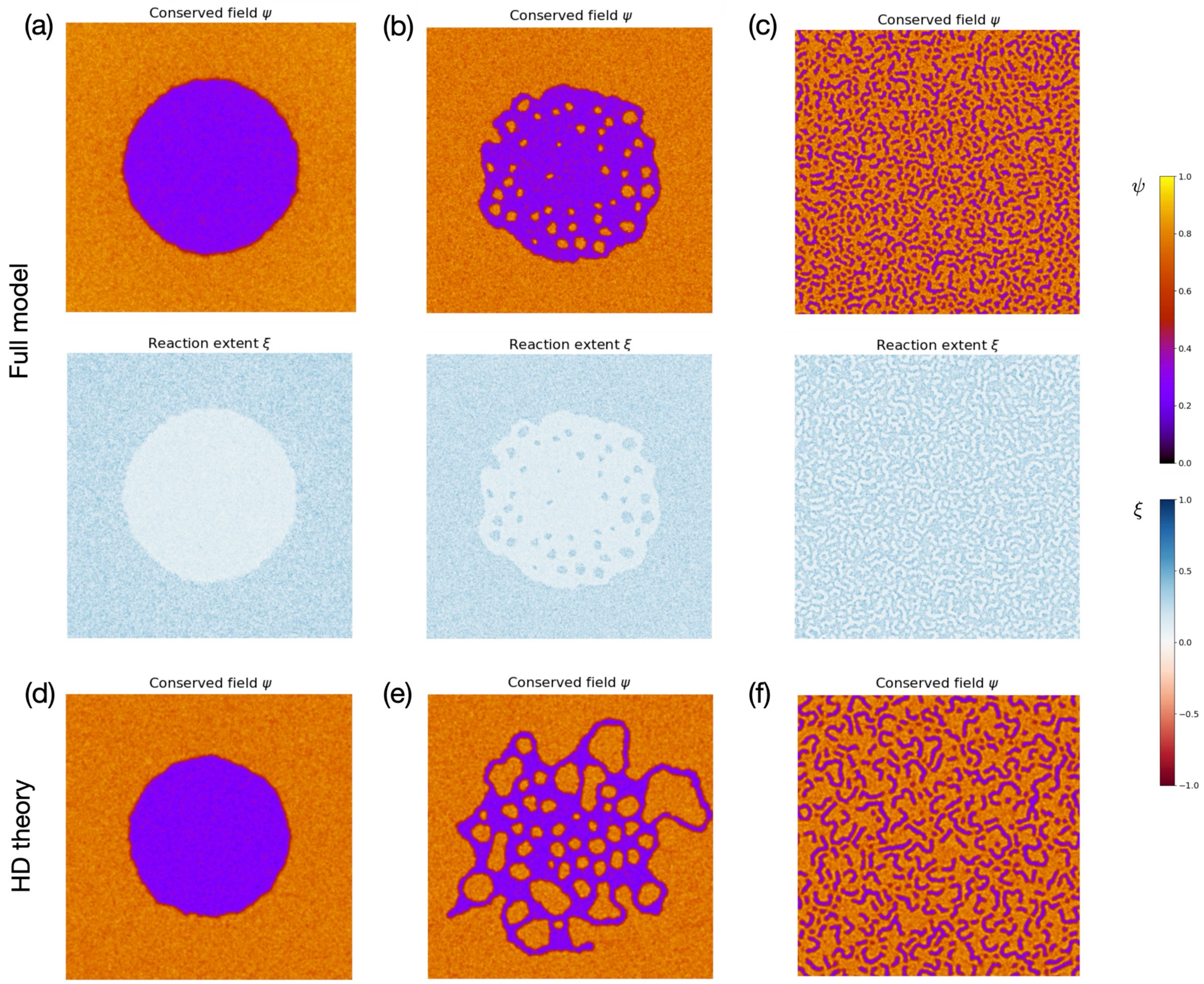}    \caption{Steady-state snapshots from simulations with noise for different $\kappa$ values and $k_m=0.558$, $\Lambda_0=1.0$, $\Delta\mu=20.0$ for $\langle \psi\rangle=0.645$. The typical simulation time is $k_mt>5\times 10^5$. For the full model we use $D=k_BT\Lambda_0v=6.25\times 10^{-3}$ and molecular volumes $v=0.02$. Different phase behaviors are for: (a) $\kappa=2.5$, leading to macrophase separation with a dilute droplet inside a dense phase, (b) $\kappa=1.8$, forming a dilute macrophase with dense microphases, (c) $\kappa=1.0$, forming an active filament phase. Corresponding simulations of the effective model in the hydrodynamic limit are with (d) $\kappa_{\rm eff}=0.47$, (e) $\kappa_{\rm eff}=-0.23$, (f) $\kappa_{\rm eff}=-1.03$. For the hydrodynamic theory (HD theory) simulations, we use $D=8.75\times10^{-3}$ (to compensate for the lack of reaction noise).}
    \label{fig:snapshots}
\end{figure*}

We examine further the arrested phases in the low density region when $\langle \psi\rangle=0.365$. We investigate how the phase behavior depends on reaction-diffusion length scales compared to interfacial length scale and driving chemical potential, i.e., on $\kappa/ (\Lambda_0/k_m)$ and $\Delta \mu$. In principle, regular arrangement of dense droplets appear due to suppression (or reversal) of Ostwald ripening \cite{zwicker2015suppression, tjhung2018cluster, hertag2026statistics}. In Fig. \ref{fig:ripening}, we show a phase diagram of ripening behavior. To determine the nature of each phase, we initiate the system with a large and a small dense droplet, and numerically solve Eqs.~\eqref{eq:freefunctional}-\eqref{eq:alpha} using the model described in this section. At long times, the system reaches one of four states: well-mixed state so no droplet, a single stable droplet via Ostwald ripening, two droplets via reverse Ostwald ripening, or multiple micro structures. The coexisting phase values of conserved field $\psi_{\rm I}$ and $\psi_{\rm II}$ also evolve with varying $\Delta\mu$. For instance, $\Delta \mu=0.5$ results $\langle \psi\rangle<\psi_{\rm I},\psi_{\rm II}$ for $\langle \psi\rangle=0.365$, hence the system remains outside of binodals and reaches a mixed state.  Using our hydrodynamic theory, we plot the line $\kappa_{\rm eff}=\kappa+K=0$ taking $K=K(\psi^{\rm int})$ at the interface value. We use the same value of $\psi^{\rm int}=0.516$ (corresponding to large $\Delta \mu$ limit) as we scan over $\Delta \mu$. Remarkably, $\kappa_{\rm eff}=0$ line appears to be an excellent prediction to distinguish the ripening behavior. Therefore, our hydrodynamic theory provides a simple quantitative framework to capture arrested microphases in this low density regime.

The same ripening diagram does not hold in the high density region for $\langle \psi\rangle$, as the phase behavior is asymmetric (Fig. \ref{fig:phasediagram}). However, as we have demonstrated, a novel phase behavior appears in this regime, i.e., ``continuously-moving filament phases" at low $\kappa$ values. In addition to what we have discussed so far, AMB+ theory suggests a micro-macro phase coexistence, namely a bubbly phase separation in the presence of noisy dynamics. Thus, we next turn our attention to stochastic simulations of the model.

 \subsection{Bubbly phase separation in the presence of noise}

So far, our hydrodynamic theory is consistent with the AMB+ model in the following way. Microphases appear when the emergent interfacial tension becomes negative due to active contributions, for which $\kappa_{\rm eff}<0$ is sufficient. The signs of $\zeta_{\rm eff}$ and $\lambda_{\rm eff}$ determine the asymmetry of the phases. If  $\zeta_{\rm eff},\lambda_{\rm eff}<0$, arrested dense droplets in a dilute background emerge in the low density regime $\langle \psi \rangle< \psi^{\rm int}$ while coarsening and dynamic phases appear in the high density regime $\langle \psi \rangle> \psi^{\rm int}$. A reverse asymmetry can be obtained under sign reversal $\zeta_{\rm eff}\rightarrow -\zeta_{\rm eff}$ and $\lambda_{\rm eff}\rightarrow -\lambda_{\rm eff}$ \cite{tjhung2018cluster}. For $\zeta_{\rm eff},\lambda_{\rm eff}<0$, the high density region phase behavior depends on both large-scale and capillary interface tensions. For $\kappa_{\rm eff}<0$, the large-scale interface tension can be negative. It has been shown for the AMB+ that when capillary tension becomes negative as well, moving phases appear due to interface instability \cite{fausti2021capillary}. In our case, we see the emergence of ``dynamic filament phases". By contrast, if capillary tension is positive, micro-macro phase co-existence can appear in systems with noisy dynamics. In this phase, a dense/dilute macrodroplet forms with continuous creation of dilute/dense microbubbles and removal of bubbles at the surface of the large droplet. In our model, we explore this possibility by varying $\kappa_{\rm eff}$ via $\kappa$ while fixing $\Delta \mu=20.0$. We also set $k_m=0.558$ and $\Lambda_0=1.0$. These result in effective parameters $K=-2.03$, $\lambda_{\rm eff}=-1.04$, $\zeta_{\rm eff}=-3.91$, $\eta_{\rm eff}=-4.35$, and $\nu_{\rm eff}=2.33$ in the hydrodynamic theory (see Section \ref{sec:5d}).

For stochastic dynamics of the full system, we include noise in both diffusive and reaction fluxes as explained in Appendix \ref{app:noise}. We take $D=k_B T\Lambda_0=6.25\times 10^{-3}$ and molecular volume of A and B as $v=0.02$. In Fig. \ref{fig:snapshots}, we show steady-state snapshots from simulations with noise for different values of $\kappa=2.5,\ 1.8,\ 1.0$ and $\langle \psi\rangle=0.645$. Remarkably, we observe micro dense droplets inside a macro dilute phase for $\kappa=1.8$. The droplets are created stochastically inside the macrophase and diffuse until ejected at the surface (videos will be available with the publication). This corresponds to the bubbly phase separation in the AMB+. $\kappa=2.5$ and $\kappa=1.0$ respectively result in a stable macrophase and continuously-moving filament phases with noise. The hydrodynamic limit of these cases gives $\kappa_{\rm eff}=0.47$, $\kappa_{\rm eff}=-0.23$, and $\kappa_{\rm eff}=-1.03$, respectively. For stochastic simulations of the hydrodynamic theory, we add noise to the flux in Eq.~\eqref{eq:dpsieff} and use a higher noise amplitude than in the full model as we did not explicitly include the noise from eliminating $\xi$ (see Appendix \ref{app:noise}). In Fig. \ref{fig:snapshots} (d), (e), (f), we show the steady-state snapshots for the corresponding $\kappa_{\rm eff}$ values. In Fig. \ref{fig:snapshots} (e), we observe that the bubbly phase separation in the hydrodynamic theory is less circular and forms much larger meso regions inside the macrophase compared to the full model. Overall, the hydrodynamic theory qualitatively captures the phase behavior for each example; however, there are differences in the characteristic length scales.

Although we considered a large $\Delta \mu$ limit, the driving chemical potential is also a determinant of active phases. We leave a comprehensive phase diagram for a future work, yet we tested different $\Delta \mu $ values when $\kappa=1.0$ and $D=6.25\times 10^{-3}$, $v=0.02$. As we have shown in Fig. \ref{fig:snapshots}(c), $\kappa=1.0$, $\Delta \mu=20.0$ results in dynamic filament phases. When $\kappa=1.0$ is kept, lowering the driving chemical potential to $\Delta\mu=2.25$ resulted in bubbly phase separation similar to the snapshot in Fig. \ref{fig:snapshots}(b) while $\Delta\mu=1.75$ resulted in a single macrophase similar to Fig. \ref{fig:snapshots}(a). Thus, both the ratio of interface and reaction-diffusion length scales $\sqrt{\kappa/ (\Lambda_0/k_m)}$ and the driving chemical potential $\Delta \mu$ are the key parameters for the phase behavior.  

Because our model is based on chemical reactions, it can provide an explanation for the microscopic mechanisms leading to the observed exotic phases. Since the chemical activity $\alpha(\psi)$ is a decreasing function of $\psi$, the production of B from A becomes more irreversible for high solvent density $1-\psi$, i.e., in dilute phase. As B accumulates in the dilute region, it phase separates with the solvent leading to nucleation of dense droplets. In the case where both dilute macrophase and dense microdroplets are stable, these dense droplets are only ejected at the surface of the macrophase. In the case where the macrophase interface becomes unstable, this leads to shape instabilities at smaller scales resulting in dynamic microphases.

\section{Entropy production rate}\label{sec:epra}
Our model is thermodynamically consistent and thus permits us to calculate the production
of entropy due to active processes driven by the non-equilibrium chemical reaction. Using irreversible thermodynamics of transport and reaction processes \cite{de2013non},  the local entropy production rate per volume can be expressed as
\begin{equation}
\dot s =\dot s_{\rm diff}+\dot s_{\rm r}\label{eq:epr}
\end{equation}
with $\dot s\geq 0$,
where
\begin{eqnarray}
T\dot s_{\rm diff}&=&-\bm{j}_{\psi}\cdot \nabla \mu_{\psi}-\bm{j}_{\xi}\cdot \nabla \mu_{\xi}\\
T\dot s_{\rm r} &=&-\sum_{i=1}^2 r_i \Delta G_i \label{eq:epr1}
\end{eqnarray}
are the contributions from diffusion and from reactions, respectively. Here,
$\Delta G_1=2\mu_{\xi}$ and  $\Delta G_2=2\mu_{\xi}-\Delta\mu$ are the Gibbs free energy of the reactions and $r_1$ and $r_2$ are the rates of the equilibrium and non-equilibrium reaction.  Note that \eqref{eq:epr} neglects the dissipation associated with heat transport that is not considered in the present work. 

\begin{figure}[h]
    \centering
    \includegraphics[width=\columnwidth]{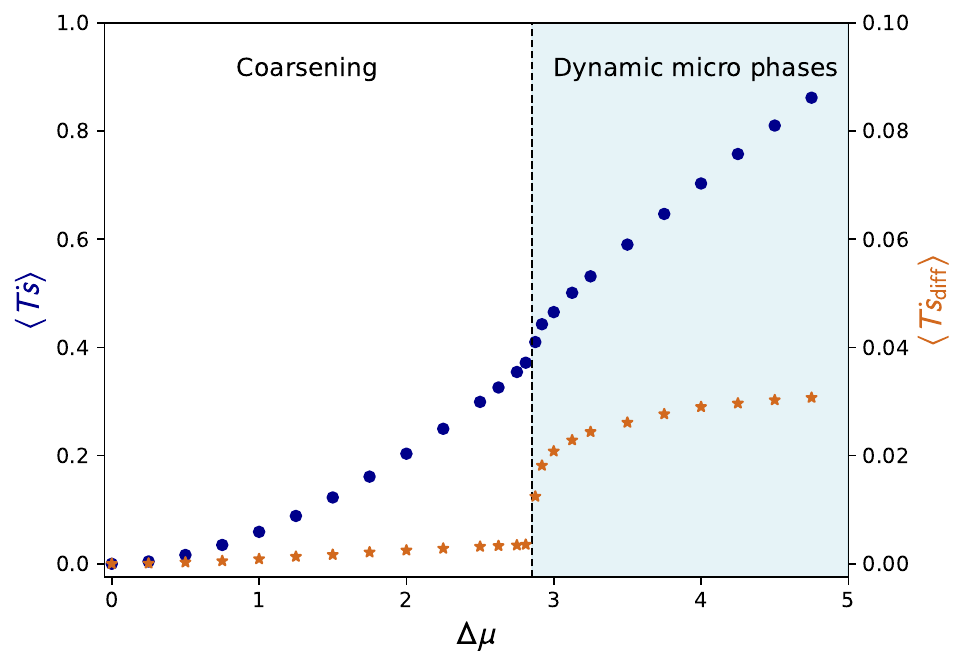}
   \caption{Entropy production rate per unit area in the system for $\langle \psi\rangle=0.65$,  $k_m=0.558$, $\Lambda_0=1.0$, and $\kappa=1.0$ in the absence of noise. The blue dots show the mean entropy production rate and orange asterisks show the diffusive contribution. Both exhibit a kink at the onset of dynamic microphases particularly visible for the diffusive contribution.}
    \label{fig:epr}
\end{figure}

We can use this  framework to study the entropy production rate for the ternary Flory-Huggins model with chemical reactions of Section \ref{sec:numerics} in two dimensions. The spatial averaged entropy production rates per unit area $\langle \dot s\rangle= \langle \dot s_{\rm diff}\rangle + \langle \dot s_{\rm r}\rangle$ is obtained from Eq.~\eqref{eq:epr1} 
with
\begin{eqnarray}
        \langle T \dot s_{\rm diff}\rangle&=&\frac{1}{A} \int T\dot s_{\rm diff}\  dA\nonumber\\
        \langle T\dot s_{\rm r}\rangle&=&\frac{1}{A} \int T \dot s_{\rm r} \ dA\label{eq:eprdenstot}
\end{eqnarray}
where $A$ is the total area of the system, $dA$ is the area element. We study the system in the absence of noise with parameters $\langle \psi\rangle=0.65$,  $\kappa=1.0$, $\Lambda=1.0$. We numerically solve Eqs.
\eqref{eq:freefunctional}-\eqref{eq:alpha} with the model of Section \ref{sec:model} and calculate the total average entropy production per area $\langle T\dot s\rangle$ and the average entropy production rate due to diffusion $\langle T\dot s_{\rm diff}\rangle$ as a function of time using  \eqref{eq:eprdenstot}. In Fig. \ref{fig:epr}, we show typical steady-state entropy production rates, which are obtained by extrapolating long time series of $\langle T \dot s \rangle$, $\langle T \dot s_{\rm diff}\rangle$ (see Appendix \ref{app:numsol}).

In a homogeneous phase, dissipation by diffusion is zero and reactions are the only contributors to $\dot{s}$. As the system phase separates, the interfaces contribute to dissipation by diffusion leading to non-zero yet still weak entropy production rate compared to the reactions for macro phase separation. However, as the dynamic phases emerge with increased $\Delta \mu$, we observe a singular behavior of the entropy production rate at the phase transition, 
see Fig. \ref{fig:epr}.

\section{Discussion and Conclusions}

In this work, we have studied theoretically active emulsions in a solution where the solute undergoes a chemical conversion reaction between two forms $\rm A$ and $\rm B$. The emulsion is active because the conversion reaction has two parallel pathways: a passive equilibrium pathway respecting detailed balance condition and an active non-equilibrium pathway, which consumes a chemostatted fuel such as ATP. The system has only one conserved parameter, the total solute concentration $\psi$, which is a slow variable that relaxes by diffusion. The other order parameter is the reaction extent, which is the difference in concentrations of the two forms $\rm A$ and $\rm B$. It has a finite relaxation time fixed by the chemical reactions and thus should be considered as a fast variable. Taking advantage of this time scale separation, we have constructed a hydrodynamic theory for the active emulsion by adiabatically eliminating the fast variable to formulate a theory valid at large length scales and long time scales for a single conserved order parameter $\psi$.
The effective theory contains time-reversal symmetry-breaking terms in the conserved field dynamics allowing for a direct mapping of the hydrodynamic theory to AMB+. However,  our theory leads to additional terms arising from the expansion in the hydrodynamic limit: one term where the flux of the conserved variable is proportional $\eta_{\rm eff}(\nabla\psi)^3$ and one where the chemical potential is proportional to $\nu_{\rm eff}\nabla^4\psi$. The latter is important when the effective interfacial coefficient $\kappa_{\rm eff}=\kappa+K$ becomes small or negative, and it is relevant for many chemical systems.

We have applied the theory to a Flory-Huggins model including chemical reactions where only one of the isoforms $\rm A$ is soluble and the other one $\rm B$ drives the local phase separation that makes the emulsion. We considered the scenario in which higher solvent concentration enhances irreversibility of production of B from A (i.e., $\beta(\psi)<0$)  and in turn accumulating B enhances phase separation. We have then shown that phase behavior depends on interfacial energy coefficient, reaction-diffusion length scales, and the driving chemical potential. We studied numerically the phase behavior both from the full original model and from the coarse-grained hydrodynamic theory.   The results of the two theories are very similar, validating the coarse-graining procedure.  An emergent negative surface tension arises due to an active tension with a negative  interfacial coefficient $K$, which is large enough that $\kappa_{\rm eff}<0$. In this range of parameters, we observe novel phase behavior of the chemically active emulsions in addition to microphase separation. We have not constructed a complete phase diagram with all the possible exotic and non-exotic phases, but we have shown examples of the exotic phase behavior: bubbly and dynamic filament phases appear in the region where the interfacial tension becomes negative enough. The dynamics of bubble phases in the full Flory-Huggins reaction model with noise are reminiscent of those arising in the AMB+: the bubbles are created inside the macrophase and migrate to the surface. The moving filament phase has resemblance with the active foam phases of the AMB+ by the continuous motion of patterns even in the absence of  noise. However, the filaments are not connected, in contrast to a foam network, therefore display a novel phase behavior. This could be relevant for modeling of organelle dynamics exhibiting growth and fission such as mitochondria \cite{viana2020mitochondrial,tabara2025molecular}.

Our theory also provides a thermodynamically consistent way of calculating the entropy production rate in the system considering the underlying physico-chemical processes. For homogeneous systems, the dissipation is only due to chemical reactions. For the phase separating systems, estimation of entropy production rate shows that the majority of the contribution still comes from the chemical reactions which are hidden in the active field theories. Moreover, our inspection shows that both the total entropy production rate and diffusive contribution increases with the driving chemical potential $\Delta \mu$ while we observe a kink in the dissipation rate due to diffusive fluxes when the dynamic phases emerge. 

Bringing our work together with the earlier literature on active field theories opens the opportunity to obtain the observed exotic phases in real-world chemical systems. In addition, our work shows that AMB+ phases are not necessarily related to any  critical phenomena and do not rely on the Landau-Ginzburg expansion of the free energy up to order $\psi^4$. They can exist more generically even far from a critical point  as in the Flory-Huggins model. Although we focused here on a choice of the active coefficient $\alpha$ that allows for explicit analytical calculations facilitating the comparison with the effective parameters of the AMB+ phases, more realistic variations of $\alpha$ with the conserved order parameters could be studied numerically. We plan an in-depth study of particular chemical systems in a future work. 

In principle, chemically active emulsions display even richer phase behavior. It will be interesting to explore whether the hydrodynamic theory in Eq.~\eqref{eq:jresult} also captures other exotic phase behaviors such as dividing droplets \cite{zwicker_growth_2016}, oscillatory phases \cite{halatek2018rethinking,brauns2024nonreciprocal}, and rotating gear phases \cite{rasshofer2025capillary}.

\acknowledgments{We acknowledge support from the Max Planck Computing and Data Facility. This work received support from the French
government under the France 2030 investment plan, as part of the Initiative d’Excellence d’Aix-Marseille Université - Amidex (AMX-23-CEI-064). It was furthermore supported by the Joachim Herz Foundation and conducted within the Max Planck School Matter to Life supported by the German Federal Ministry of Education and Research (BMBF) in collaboration with the Max Planck Society.}

\appendix
\section{Linear transformations of density field variables} \label{app:lineartransform}
We have defined the conserved density and the reaction extent as $\psi=\phi_{\rm A}+\phi_{\rm B}$ and $\xi=\phi_{\rm B}-\phi_{\rm A}$. Then, the variables 
$\phi_A$ and $\phi_B$ 
can be transformed to $\psi$ and $\xi$ using a transformation matrix:
\begin{equation}
\mathbb{T}=\left(
\begin{array}{cc}
1  & 1  \\
-1 & 1 
\end{array}
\right) \label{eq:tmatrix}
\end{equation}
which satisfies:
\begin{equation}
\left(
\begin{array}{c}
\psi \\ \xi
\end{array}
\right)=\mathbb{T}\left(
\begin{array}{c}
\phi_{\rm A} \\ \phi_{\rm B} 
\end{array}
\right) \label{eq:trpsi} \quad .
\end{equation}
Accordingly, the following transformation also hold for the mobility matrix:
\begin{equation}
  \left(
\begin{array}{c c}
\Lambda_{\psi\psi} & \Lambda_{\psi\xi} \\ \Lambda_{\xi\psi} & \Lambda_{\xi\xi}
\end{array}
\right)=\mathbb{T}\left(
\begin{array}{c c}
\Lambda_{\rm AA} &  \Lambda_{\rm AB}\\ \Lambda_{\rm BA} & \Lambda_{\rm BB}
\end{array}
\right)\mathbb{T}^T\label{eq:trmob} 
\end{equation}
where ${\mathbb{T}^{T}}$ is the transpose of $\mathbb{T}$. This form conserves positive definiteness under linear transformation and leads to $\Lambda_{\psi\psi}=\Lambda_{\rm AA}+2\Lambda_{\rm AB}+\Lambda_{\rm BB}$, $\Lambda_{\psi\xi}=\Lambda_{\xi\psi}=\Lambda_{\rm BB}-\Lambda_{\rm AA}$, and $\Lambda_{\xi\xi}=\Lambda_{\rm AA}-2\Lambda_{\rm AB}+\Lambda_{\rm BB}$ where we used $\Lambda_{\rm AB}=\Lambda_{\rm BA}$. Finally, chemical potentials are related by:
\begin{equation}
 \left(
\begin{array}{c}
\mu_\psi \\ \mu_\xi
\end{array}
\right)=({\mathbb{T}^{T}}) ^{-1}\left(
\begin{array}{c}
\mu_{\rm A} \\ \mu_{\rm B} 
\end{array}
\right)  \label{eq:trmus} \quad 
\end{equation}
 where $({\mathbb{T}^{T}}) ^{-1}$ is the inverse of the transpose matrix. This leads to $\mu_{\psi}=(\mu_{\rm A}+\mu_{\rm B})/2$ and $\mu_{\xi}=(\mu_{\rm B}-\mu_{\rm A})/2$.

 \begin{figure}[h]
    \centering
    \includegraphics[width=\columnwidth]{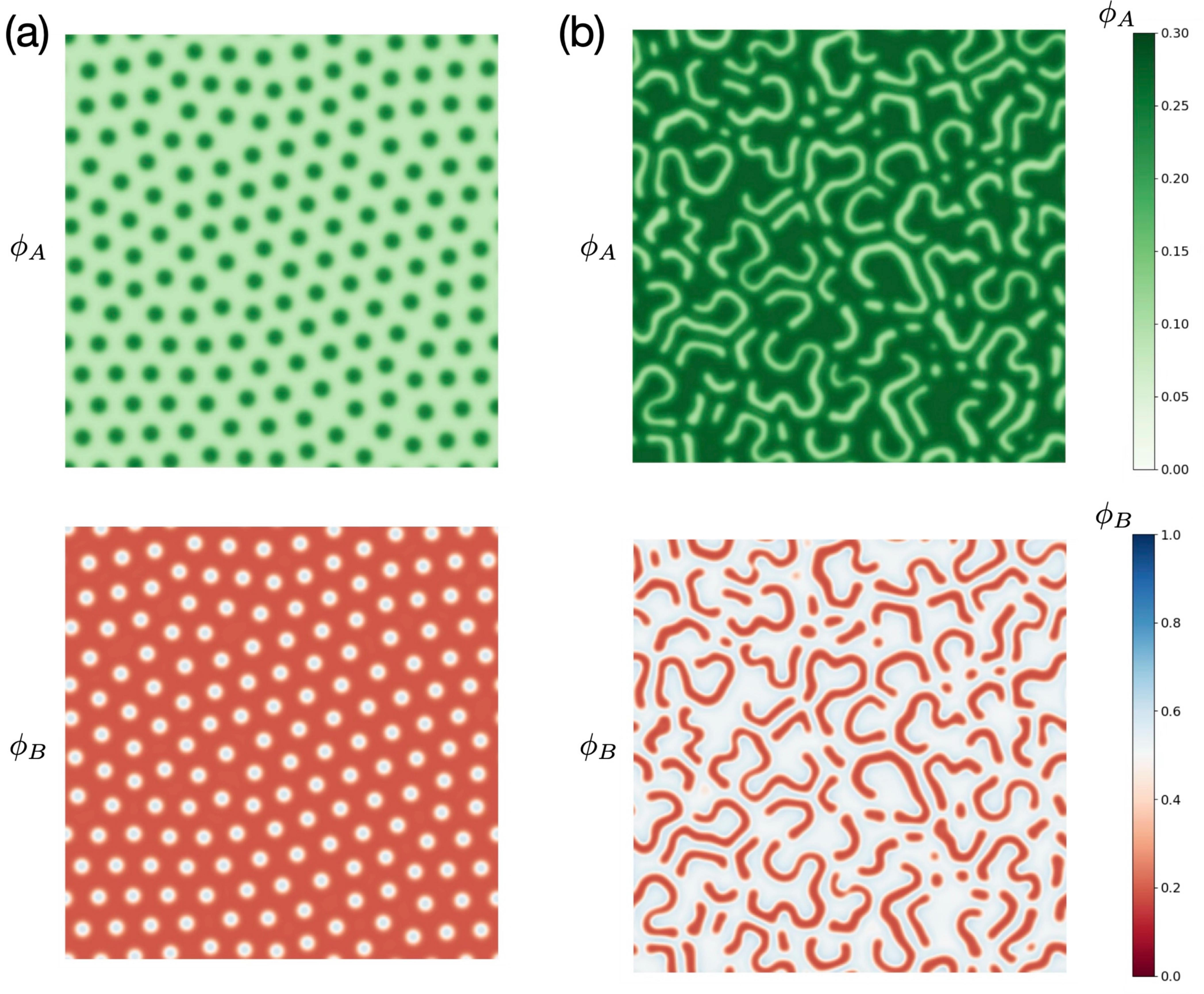}
   \caption{Snapshots of volume fractions $\phi_A$ and $\phi_B$ corresponding to (a) Fig.\ref{fig:phasediagram}(a) and (b) Fig.\ref{fig:phasediagram}(d).}
    \label{fig:ab}
\end{figure}

\section{Systematic hydrodynamic expansion} \label{app:expansion}

In this section, we present the details of the calculation outlined in section~\ref{sec:hydro}, arriving at the result presented in section~\ref{sec:activeB}.
To lowest order, we have already found $\xi_0=g(\psi)$ defined via Eq.~\eqref{eq:muxi0}. From $\mu_\xi^{(0)}(\psi,g(\psi))
 =-\frac{1}{2}\ln(1-\alpha(\psi,g(\psi)))$, the function $g$ must obey 
 $(\partial\mu_\xi^{(0)}/\partial\psi) +(\partial\mu_\xi^{(0)}/\partial\xi) g'=\frac1{2(1-\alpha)} ((\partial\alpha/\partial\psi) + (\partial\alpha/\partial\xi) g')$ at $\xi=g(\psi)$ and thus $g'$ is given as in Eq.~\eqref{eq:gprime}.

We now calculate the first order in $\epsilon$ as described in the main text. To that end we rewrite in terms of the effective chemical potential defined in Eq.~\eqref{eq:mueff} to lowest order, which is
\begin{equation}
\mu_{\rm eff}^{(0)}=\mu_{\psi}^{(0)}(\psi,g(\psi))+ \int_0^\psi \mathrm{d}u  \frac{\mathrm{d}\mu_{\xi}^{(0)} (u,g(u))}{\mathrm{d}u} \frac{\Lambda_{\psi\xi}(u,g(u))}{\Lambda_{\psi\psi}(u,g(u))} .
\end{equation}
With this,
\begin{eqnarray}
{\bm j}_\psi^{(0)}&=&-\Lambda_{\psi\psi}\nabla_{\rho}\mu_{\rm eff}^{(0)} , \label{eq:jpsi0}\\
{\bm j}_\xi^{(0)} &=& -\Lambda_{\psi\xi}\nabla_{\rho}\mu_{\rm eff}^{(0)} -\Gamma \nabla_{\rho} \mu_{\xi}^{(0)} 
\end{eqnarray}
where $
\Gamma=\Lambda_{\xi\xi}-\frac{\Lambda_{\psi\xi}^2}{\Lambda_{\psi\psi}}$.
From setting $\partial \psi/\partial \tau=-\bar{D}^{-1}\nabla_{\rho}\cdot{\bm j}_\psi^{(0)}$ in Eq.~\eqref{eq:Oepsilon}, we determine
\begin{multline}
  \mu_\xi^{(1)} = \frac{\bar{k}}{4\bar{D}k}
  \left[ -\left(g'\nabla_{\rho}\cdot\Lambda_{\psi\psi}
  - \nabla_{\rho}\cdot\Lambda_{\psi\xi}\right)\nabla_{\rho}\mu_{\rm eff}^{(0)}
  \right. \\ \left.
  + \nabla_{\rho}\cdot\Gamma\,\nabla_{\rho}\mu_{\xi}^{(0)} \right]
  \label{eq:muxi1rho}
\end{multline}
in units $(\rho,\tau)$. In this and the following results, the right hand side is evaluated at $\xi=g(\psi)$. From Eq.~\eqref{eq:muxi1rho}, we obtain in original units $(x,t)$ for $\mu_\xi=\mu_\xi^{(0)}+\epsilon \mu_\xi^{(1)}$
\begin{equation}
\epsilon\mu_\xi^{(1)}= \hat A \nabla \mu_{\rm eff}^{(0)}+\frac{1}{4k}\nabla \cdot \Gamma \nabla \mu_{\xi}^{(0)}.\label{eq:muxi1}
\end{equation}
where we have defined the operator
$\hat A=(\nabla \Lambda_{\psi\xi}-
g'\nabla \Lambda_{\psi\psi})/4k$.
We also calculate the correction to $\xi=\xi_0+\epsilon\xi_1$, using $\epsilon \mu_\xi^{(1)}=\mu_\xi(\psi, \xi_0+\epsilon \xi_1)-\mu^{(0)}_\xi(\psi, \xi_0+\epsilon \xi_1)$, which implies
\begin{equation}
\epsilon\mu_{\xi}^{(1)}=\frac{\partial\mu_\xi}{\partial {\xi}}\epsilon\xi_1+\frac{1}{2}\frac{\partial \ln(1-\alpha)}{\partial \xi}\epsilon\xi_1\quad .
\end{equation}
Therefore, we have
\begin{eqnarray}
\xi_1&=&h\mu_{\xi}^{(1)}, \quad h=\frac{1}{\chi^{-1}_{\xi \xi}-\frac{1}{2}\frac{\partial \alpha}{\partial \xi} /(1-\alpha)}
\quad .\label{eq:xi1}
\end{eqnarray}
This provides the correction to order $\epsilon$ to $\xi(\psi)$. Finally, we can determine the contribution $\epsilon\mu_{\psi}^{(1)}$ to $\mu_\psi$ as
\begin{eqnarray}
 \epsilon\mu_{\psi}^{(1)}&=&\frac{\partial \mu_{\psi}}{\partial \xi}  \epsilon\xi_1-\kappa\nabla^2\psi\\
 &=&(-g'+\gamma h)\epsilon\mu_{\xi}^{(1)}-\kappa\nabla^2\psi \label{eq:mupsi1} \quad ,
\end{eqnarray}
where we have defined 
\begin{equation}
\gamma=\frac{\partial \mu_\xi^{(0)}}{\partial \psi} 
= \frac{1}{2}\frac{\partial\alpha/\partial\psi}{1-\alpha}
\end{equation}
and used $\chi^{-1}_{\psi\xi}h=g'-\gamma h$. Next, we insert $\epsilon\mu_{\xi}^{(1)}$ from Eq.~\eqref{eq:muxi1}, leading to:
\begin{equation}
\epsilon\mu_{\psi}^{(1)}=-\left(g'-\gamma h\right) \left(\hat A\nabla \mu_{\rm eff}^{(0)} +\frac{1}{4k}\nabla \cdot \Gamma \nabla \mu_{\xi}^{(0)}\right) 
-\kappa\nabla^2\psi \ .\label{eq:mupsi1_v3}
\end{equation}

Now we can calculate the dynamics $\partial \psi/ \partial t = -\nabla \cdot {\bm j}$ of the conserved density. By expanding also  $s(\psi,\xi_0+\epsilon \xi_1)=s^{(0)}+\epsilon s^{(1)}$, the flux $\bm{j}$ from Eq.~\eqref{eq:jpsi} can be rewritten as
\begin{equation}
    {\bm j}=-\Lambda_{\psi\psi}\left(\nabla(\mu_{\rm eff}^{(0)}+\epsilon\mu_{\psi}^{(1)})+s^{(0)}\nabla\epsilon\mu_{\xi}^{(1)}+\epsilon s^{(1)}\nabla\mu_{\xi}^{(0)}\right) \quad. \label{eq:japp}
\end{equation} 
Here $s^{(0)}=s(\psi,g(\psi))$ and $s^{(1)}=(\partial s/\partial \xi)\xi_1$.
We can rewrite the last term in Eq.\eqref{eq:japp} as 
\begin{eqnarray}
     \epsilon s^{(1)}\nabla\mu_{\xi}^{(0)}&=& \epsilon s_{\xi} \xi_1 \nabla\mu_{\xi}^{(0)}\nonumber\\  &=& \epsilon s_{\xi}h\mu_{\xi}^{(1)} \nabla\mu_{\xi}^{(0)}\label{eq:es1}\quad ,
\end{eqnarray}
where $s_{\xi}=\partial s/\partial \xi$, and we used Eq. \eqref{eq:xi1} in the second line. Using Eq. \eqref{eq:muxi1} we write 
\begin{eqnarray}
      \epsilon s_\xi h \mu_\xi^{(1)}\nabla\mu_{\xi}^{(0)}
      &= &\nabla\mu_{\xi}^{(0)} s_{\xi} h \hat A \nabla\mu_{\rm eff}^{(0)} - \mu_{\xi}^{(0)}\nabla  s_{\xi} h \hat A    \nabla\mu_{\rm eff}^{(0)} \nonumber\\
      &+&\frac{s_{\xi}h}{4k}(\nabla \cdot \Gamma \nabla \mu_{\xi}^{(0)}) \nabla \mu_{\xi}^{(0)}\label{eq:s1term}
\end{eqnarray}

Finally, inserting Eqs.~\eqref{eq:muxi1}, \eqref{eq:mupsi1_v3}, and \eqref{eq:s1term} in \eqref{eq:japp} we obtain
\begin{eqnarray}
{\bm j}&=&-\hat M \nabla \mu_{\rm eff}^{(0)} \nonumber\\
&&-
\Lambda_{\psi\psi}\nabla \left(-\kappa\nabla^2\psi -\frac{g'-\gamma h}{4k}\nabla \cdot \Gamma\nabla \mu_{\xi}^{(0)}\right)
\nonumber\\
&&-
\Lambda_{\psi\psi}s\nabla \frac{1}{4k}\nabla \cdot \Gamma \nabla\mu_{\xi}^{(0)}\nonumber\\
&&-\Lambda_{\psi\psi}\frac{s_{\xi}h}{4k}(\nabla \cdot \Gamma \nabla \mu_{\xi}^{(0)} ) \nabla \mu_{\xi}^{(0)}\label{eq:jeff}
\end{eqnarray}
where the operator $\hat M$ is given in Eq.~\eqref{eq:M}. We can write the mobility operator $\hat M$ in front of all terms. This rearrangement only generates terms of higher order in $\epsilon$, which we can drop in the
hydrodynamic limit. We thus have:
\begin{eqnarray}
{\bm j}&=&-\hat M \nabla \left(\tilde{\mu}_{\rm eff} - \frac{g'-\gamma h}{4k}\nabla \cdot \Gamma\nabla \mu_{\xi}^{(0)} \right)
\nonumber\\
&&-
\hat M s \nabla \frac{1}{4k}\nabla \cdot \Gamma \nabla\mu_{\xi}^{(0)}
\nonumber\\
&&-
\hat M \frac{s_{\xi}h}{4k}(\nabla \cdot \Gamma \nabla \mu_{\xi}^{(0)} )\nabla \mu_{\xi}^{(0)} \quad \label{eq:jeff_v2}
\end{eqnarray}
with 
\begin{equation}
\tilde{\mu}_{\rm eff}=\mu_{\rm eff}^{(0)}-\kappa\nabla^2\psi .
\end{equation}
We rearrange further to the form
\begin{equation}
\hat M^{-1} {\bm j}=-\nabla(\tilde{\mu}_{\rm eff} +{\tilde\mu})-\bf{C} 
\end{equation}
with
\begin{eqnarray}
{\tilde\mu}&=& -(g'-\gamma h-s)\frac{1}{4k}\nabla \cdot \Gamma \nabla
\mu_{\xi}^{(0)}
 \\
\bf{C}&=&-(\nabla s ) \left(\frac{1}{4k}\nabla \cdot \Gamma \nabla\mu_{\xi}^{(0)}\right) + \frac{s_{\xi}h}{4k}(\nabla \cdot \Gamma \nabla \mu_{\xi}^{(0)}) \nabla \mu_{\xi}^{(0)} \nonumber .
\end{eqnarray} 
Noting that
\begin{eqnarray}
\nabla\mu_{\xi}^{(0)}&=&\beta\nabla\psi , \\
\nabla^2\mu_{\xi}^{(0)}&=&\beta\nabla^2\psi+\beta'(\nabla\psi)^2 
\end{eqnarray}
where $\beta \equiv d \mu_\xi^{(0)}/d \psi  = \frac{1}{2}\frac{\alpha'}{1-\alpha} $ and $\beta'=d\beta/d\psi$, we finally obtain
\begin{align}
  \tilde{\mu} &= -\frac{g'-\gamma h-s}{4k}
  \left[\Gamma\beta\nabla^2\psi
  + \Gamma\beta'(\nabla\psi)^2
  + \Gamma'\beta(\nabla\psi)^2\right] , \nonumber\\[6pt]
  \mathbf{C} &= \frac{s_{\xi}\beta h - s'}{4k}
  \left[\Gamma'\beta(\nabla\psi)^2
  + \Gamma\beta\nabla^2\psi \right. \nonumber\\ 
  &\qquad \hspace{8em} \left.
  + \Gamma\beta'(\nabla\psi)^2\right]\nabla\psi \ . 
\end{align}
 Collecting the terms in $(\nabla\psi)^2$, $\nabla^2 \psi$, $(\nabla^2 \psi)(\nabla\psi)$
 and $(\nabla \psi)^3$, we obtain Eqs. \eqref{eq:jresult}--\eqref{eq:M}.\\

\section{Stabilizing higher order gradient term}\label{app:q6}
Here we show the derivation of the coefficient $\nu_{\rm eff}$ that appear in \eqref{eq:jresult}. This term becomes important when $\kappa_{\rm eff}$ is small or negative. This is the only term contributing to linear stability at next order, $\mathcal{O}(\epsilon^2)$ in the hydrodynamic expansion. We can take a shortcut calculation ignoring all non-linear terms at order $\epsilon^2$. Our aim is to write the dynamics in Eq. \eqref{eq:dynamics} at next order in $\epsilon$ in the spirit of Eqs.\eqref{eq:xieps} and \eqref{eq:psips}. Now, we can write in original units the next order in $\epsilon$ expansion as:
\begin{widetext}
\begin{align}\label{eq:eps2}
   - \epsilon^2 g'\nabla\cdot\bm{j}_{\psi}^{(1)}-\epsilon^2\left(\frac{\partial\xi_1}{\partial \psi}+\frac{\partial\xi_1}{\partial \nabla\psi}\cdot{\nabla}+\frac{\partial\xi_1}{\partial\nabla^2 \psi}\nabla^2\right)\nabla\cdot\bm{j}_{\psi}^{(0)} =-\epsilon^2 \nabla \cdot \bm{j}_\xi^{(1)} -4k^{(0)}\epsilon^2 \mu_{\xi}^{(2)} -2k^{(0)}\epsilon^2{\mu_{\xi}^{(1)}}^2 -4k^{(1)}\epsilon^2{\mu_{\xi}^{(1)}}  
\end{align}
\end{widetext}
using $k=k^{(0)}+\epsilon k^{(1)}$, ${\bm j}_m={\bm j}_m^{(0)}+\epsilon{\bm j}_m^{(1)}$ for $m\in \{\psi,\xi \}$ and $\partial(\nabla\psi)/\partial t=\nabla 
(\partial\psi/\partial t)$, $\partial(\nabla^2\psi)/\partial t=\nabla^2 (\partial\psi/ \partial t)$. Eq.\eqref{eq:eps2} can be solved for $\mu_{\xi}^{(2)}$. We 
are only interested in terms at $\nabla^2\psi$ from order $\epsilon$ that can  produce $\mathcal{O}(\nabla^4 \psi)$ terms in effective chemical potential for 
hydrodynamic limit of $\psi$ evolution. Thus, we can ignore all non-linear combinations of first order terms such as  ${\mu_{\xi}^{(1)}}^2$ and $k^{(1)}
{\mu_{\xi}^{(1)}}
$. Similarly, we ignore $(\nabla\psi)^2, (\nabla\psi)^3$ terms. Then, we notice that all the remaining terms in left hand side enter to 
mobility by being coupled to $\nabla \mu_{\rm eff}$ given in Eq.\eqref{eq:mueff} using \eqref{eq:jeff} for $\bm{j}_{\psi}^{(1)}$ and knowing $\bm{j}_{\psi}^{(0)}$ from \eqref{eq:jpsi0}. As a result, we can write using Eq.\eqref{eq:eps2}:
\begin{equation}
   \epsilon^2 \mu_{\xi}^{(2)}=-\frac{\epsilon^2}{4k^{(0)}}\nabla \cdot \bm{j}_\xi^{(1)}+\text{ other terms .} \label{eq:muxi2}
\end{equation}
By using \eqref{eq:js}, we have 
\begin{equation}
\bm{j}_{\xi}=s\bm{j}_{\psi}-\Gamma\nabla\mu_{\xi}
\end{equation}
which holds true irrespective of expansion order. Then, we first write $\bm{j}_{\xi}^{(1)}=s^{(0)}\bm{j}_{\psi}^{(1)}+s^{(1)}\bm{j}_{\psi}^{(0)}+\Gamma^{(0)}\nabla\mu_{\xi}^{(1)}+\Gamma^{(1)}\nabla\mu_{\xi}^{(0)}$ and apply the same procedure as above. All $\nabla^2\psi$ terms under $s^{(0)} \bm{j}_{\psi}^{(1)}$ also enter mobility-coupling integrable terms while $s^{1}\bm{j}_{\psi}^{(0)}$ and $\Gamma^{(1)}\nabla\mu_{\xi}^{(0)}$ can not produce a $\nabla^2\psi$ term. Separating out these terms, we can write \eqref{eq:muxi2} as:
\begin{equation}
\epsilon^2\mu_\xi^{(2)}= \frac{\epsilon^2}{4k^{(0)}}\nabla \Gamma^{0}\nabla\mu_{\xi}^{(1)}+ \text{other terms} \quad .
\end{equation}
We can now write explicitly the relevant $\nabla^4\psi$ order terms by inserting $\mu_{\xi}^{(1)}$ from \eqref{eq:muxi1} and ignore $\nabla\mu_{\rm eff}$ terms that enter in mobility at $\epsilon^2$ order. As a result, we have
\begin{equation}
\epsilon^2\mu_\xi^{(2)}= \epsilon^2\left(\frac{\Gamma}{4k}\right)^2\beta\nabla^4 \psi+ \text{other terms} \quad .\label{eq:muxi2v2}
\end{equation}
where we dropped the superscript (0) for brevity. Now we calculate the correction to $\xi=\xi_0+\epsilon\xi_1+\epsilon^2\xi_{2}$, using $\epsilon \mu_\xi^{(1)}+\epsilon^2 \mu_\xi^{(2)}=\mu_\xi(\psi, \xi_0+\epsilon \xi_1+\epsilon^2\xi_{2})-\mu^{(0)}_\xi(\psi, \xi_0+\epsilon \xi_1+\epsilon^2\xi_{2})$  which implies
\begin{eqnarray}
\epsilon^2\mu_{\xi}^{(2)}=&&\left(\frac{\partial\mu_\xi}{\partial {\xi}}+\frac{1}{2}\frac{\partial \ln(1-\alpha)}{\partial \xi}\right)\epsilon^2\xi_2 \nonumber \\&&+ \left(\frac{\partial^2\mu_\xi}{\partial {\xi}^2}+\frac{1}{2}\frac{\partial^2 \ln(1-\alpha)}{\partial \xi^2}\right) \frac{\epsilon^2\xi_1^2}{2} \quad.
\end{eqnarray}
As before, we simply ignore square terms $\epsilon^2\xi_1^2$ that can not generate $\nabla^4\psi$ terms. Therefore, we can write 
\begin{eqnarray}
\xi_2&=&h\mu_{\xi}^{(2)}+\text{other terms}
\quad .\label{eq:xi2}
\end{eqnarray}
Finally, we need to insert \eqref{eq:muxi2v2} and \eqref{eq:xi2} in ${\bm j}_{\psi}={\bm j}_{\psi}^{(0)}+\epsilon {\bm j}_{\psi}^{(1)}+\epsilon^2{\bm j}_{\psi}^{(2)}$ where 
\begin{multline}
  \bm{j}_{\psi}^{(2)} = -\Lambda_{\psi\psi}
  \Bigl(\nabla\mu_{\psi}^{(2)} + s^{(0)}\nabla\mu_{\xi}^{(2)} \\
  + s^{(1)}\nabla\mu_{\xi}^{(1)} + s^{(2)}\nabla\mu_{\xi}^{(0)}\Bigr)
\end{multline}
and $\mu_{\psi}^{(2)}=({\partial \mu_\xi^{(0)}}/{\partial \psi}) \xi_2$ $+$ terms with $\xi_1^2$ (which we ignore). Among all, $\mu_{\psi}^{(2)}=({\partial \mu_\xi^{(0)}}/{\partial \psi}) \xi_2$ and $s^{(0)}\nabla \mu_{\xi}^{(2)}$ are the only terms that can generate $\nabla^4\psi$ term in the chemical potential. We can write 
\begin{equation}
\nabla  ({\partial \mu_\psi^{(0)}}/{\partial \psi}) \xi_2+s^{(0)}\nabla \mu_{\xi}^{(2)} =\nabla (-g+\gamma h)\mu_{\xi}^{(2)}+s\nabla \mu_{\xi}^{(2)}  
\end{equation}
where the coefficients are evaluated at $\xi=g(\psi)$. This yields a term $\nu_{\rm eff}\nabla^4\psi$ by inserting \eqref{eq:muxi2v2}
\begin{equation}
    \nu_{\rm eff}=\left(\frac{\Gamma}{4k}\right)^2\beta
\left (-g'+\gamma h+s\right )=-\frac{\Gamma}{4k}K \ \label{eq:nueff0}
\end{equation}
which is Eq.\eqref{eq:nueff}.

\section{Stochastic simulations with noise}
\label{app:noise}

The stochastic diffusive fluxes become:
\begin{eqnarray}
    \bm{j}_\psi^{\rm st}&=&\bm{j}_\psi+\bm{\Omega}_{\psi}\quad ,\nonumber\\
    \bm{j}_\xi^{\rm st}&=&\bm{j}_\xi+\bm{\Omega}_{\xi} \label{eq:jsstch}
\end{eqnarray}
where $j_{\psi}$, $j_{\xi}$ are given by \eqref{eq:js} and $\bm{\Omega}_{\psi}$, $\bm{\Omega}_{\xi}$ are the noise terms which  should satisfy cross correlations given by the mobility matrix such that 
\begin{eqnarray}
\langle\Omega_{i}^{\psi}(\mathbf{x},t) \Omega_{j}^{\psi}(\mathbf{x}',t')\rangle&=&2k_B T\Lambda_{\psi\psi}v\delta_{ij}\delta(\mathbf{x}-\mathbf{x}')\,\delta(t-t')\ , \nonumber\\ \langle\Omega_{i}^{\psi}(\mathbf{x},t) \Omega_{j}^{\xi}(\mathbf{x}',t')\rangle&=&2k_B T\Lambda_{\psi\xi}v\delta_{ij}\delta(\mathbf{x}-\mathbf{x}')\,\delta(t-t')\ ,\nonumber\\ \langle\Omega_{i}^{\xi}(\mathbf{x},t) \Omega_{j}^{\xi}(\mathbf{x}',t')\rangle&=&2k_B T\Lambda_{\xi\xi}v\delta_{ij}\delta(\mathbf{x}-\mathbf{x}')\,\delta(t-t') \nonumber\\ &&
\end{eqnarray}
where the components of the mobility matrix are 
$\Lambda_{\psi\psi}={\Lambda_0}(1-\psi)\psi$, $\Lambda_{\psi\xi}=\Lambda_{\xi\psi}={\Lambda_0}(1-\psi)\xi$, 
$\Lambda_{\xi\xi}={\Lambda_0}(\psi-\xi^2)$. We make the following ansatz: $\bm{\Omega}_{\psi}=\sigma_{11}\bm{\Xi}_1+\sigma_{12}\bm{\Xi}_2$, $\bm{\Omega}_{\xi}=\sigma_{21}\bm{\Xi}_1+\sigma_{22}\bm{\Xi}_2$. Here, the two independent Gaussian white noise terms satisfy 
\begin{eqnarray}
\langle \Xi^i_{m}(\mathbf{x},t)\rangle&=&0 \quad ,\nonumber\\
\langle \Xi^i_{m}(\mathbf{x},t)\,\Xi^j_{k}(\mathbf{x}',t') \rangle
&=&\,
\delta_{mk}\delta_{ij}\,\delta(\mathbf{x}-\mathbf{x}')\,\delta(t-t')\nonumber\\ &&\label{eq:gaussians}
\end{eqnarray}
where $i,j$ represent spatial coordinates and $m,k\in \{1,2\}$.  The following choice satisfies these conditions with real coefficients in the physical domain ($0<\psi<1$):
\begin{eqnarray}
\sigma_{11}&=&\left(2D\psi(1-\psi)\right)^{1/2} \quad ,\nonumber\\
\sigma_{12}&=&0 \quad ,\nonumber\\
\sigma_{21}&=&\left(2D\frac{(1-\psi)\xi^2}{\psi}\right)^{1/2}\quad ,\nonumber\\
\sigma_{22}&=&\left(2D\frac{(\psi^2-\xi^2)}{\psi}\right)^{1/2}\quad\label{eq:signoise} 
\end{eqnarray}
where $D=k_B T \Lambda_0 v$. 

Moreover, we add a stochastic term in the reaction flux by using 
\begin{equation}
    r^{\rm st}=r+\Omega_r \label{eq:rstch}
\end{equation}
where $r$ is given by \eqref{eq:r} and $\Omega_r$ is the noise term. We assume that the active and passive reactions are statistically independent. Then, the noise term should satisfy:
\begin{equation}
    \langle \Omega_r(\mathbf{x},t)\,\Omega_r(\mathbf{x}',t') \rangle
=v(r_1^+ + r_2^+ + r_1^- + r_2^-)\delta(\mathbf{x}-\mathbf{x}')\delta(t-t')\ . \label{eq:rnoise}
\end{equation}
The noise amplitude is obtained by summing the contributions of the forward and backward reaction fluxes. We use a chemical Langevin description \cite{gillespie2000chemical} such that the noise is Gaussian with the form $\Omega_r=\sigma_r \Xi_r$ where $\Xi_r$ is a Gaussian white noise satisfying $\langle \Xi_r(\mathbf{x},t)\rangle=0$ and $\langle \Xi_r(\mathbf{x},t)\,\Xi_r(\mathbf{x}',t') \rangle
=\delta(\mathbf{x}-\mathbf{x}')\delta(t-t')$. Then, implementing the condition \eqref{eq:rnoise} with values of reaction fluxes from \eqref{eq:r}, we get:
\begin{equation}
    \sigma_r=(kv(1+(1-\alpha)e^{2\mu_{\xi}}))^{1/2}\quad .
\end{equation}

For the stochastic dynamics of the hydrodynamic theory, we use:
\begin{eqnarray}
    \bm{j}^{\rm st}&=&\bm{j}+\bm{\Omega}\quad \label{eq:jst}
\end{eqnarray}
where $\bm{j}$ is given by \eqref{eq:jresult} and $\bm{\Omega}=\sigma_{11}\bm{\Xi}$ with the same $\sigma_{11}$ in \eqref{eq:signoise}. The Gaussian white noise term satisfy the same conditions of \eqref{eq:gaussians}. Note that \eqref{eq:jst} only includes the noise that stems from $\bm{j}_\psi^{\rm st}$ at lowest order in the hydrodynamic expansion and we did not explicitly calculate higher order contributions as well as the noise coming from elimination of $\xi$. Instead, we use a higher $D$ value in the hydrodynamic theory compared to the full model.

 \section{Numerical methods} \label{app:numsol}
 For the numerical solutions, we have used a pseudospectral method on a regular lattice with size $L_x\times L_y$ and periodic boundary conditions. We discretize space as $x=i\Delta x$, $y=j\Delta y$ where $0\leq i\leq L_x/\Delta x$ and $0\leq j\leq L_y/\Delta y$. We use $\Delta x=1$, $\Delta y=1$. The time integration of deterministic models are performed by using an implicit-explicit third-order Runge-Kutta scheme \cite{ascher1997implicit}. The time integration of stochastic simulations are performed by using implicit–explicit Euler–Maruyama method \cite{kloedenplaten1992}. The typical time step sizes are $\Delta t=0.08$ for the deterministic and $\Delta t=0.01$ for the stochastic simulations. The lattice sizes are respectively $L_x=L_y=256$ in Fig. \ref{fig:phasediagram}, $L_x=128$, $L_y=64$ in Fig. \ref{fig:ripening}, $L_x=L_y=512$ in Fig. \ref{fig:snapshots}, and $L_x=L_y=128$ in Fig. \ref{fig:epr}.\\

 \section{Calculation of mean wave number} 
 \label{app:structure}
 We define $\delta \psi({\bf r}_n)=\psi({\bf r}_n)-\langle \psi\rangle$ at lattice point ${\bf r}_n$ where $\langle \psi\rangle$ is the average value of the conserved order parameter. Then, we calculate the Fourier transform $   \delta \tilde{\psi}(\mathbf{q}) = \sum_{n} \delta \psi(\mathbf{r}_n)\, 
e^{i \mathbf{q} \cdot \mathbf{r}_n}\, \Delta x \, \Delta y,$ which is a discretized approximation of a continuous Fourier transform over lattice sites labeled by $n$ where ${\bf q}$ is the wave vector. The structure factor is the power spectrum of $\delta\tilde{\psi}({\bf q})$, and $S({\bf q)}=A^{-1} |\delta\tilde{\psi}({\bf q})|^2$ where area $A=L_xL_y$. Using this we can obtain $S(q)$ by binning the magnitude $q=|{\bf q}|$ of wave number. Then, we calculate numerically the mean wave number by using:
\begin{equation}
    q^* = \frac{\int_0 ^{\infty}q\  S(q)  dq}{\int_0 ^{\infty}S(q) dq}\quad .
\end{equation}

\bibliography{main}
\end{document}